\documentclass[acmtog,journal]{acmart}

\newcommand{\secref}[1]{Section \ref{sec:#1}}
\newcommand{\appref}[1]{Appendix \ref{app:#1}}
\newcommand{\figref}[1]{Figure \ref{fig:#1}}
\newcommand{\equref}[1]{(\ref{eq:#1})}

\newcommand{\figloc}[1]{\emph{(#1)}}
\newcommand{\ie}{i.e.}

\usepackage{afterpage}
\usepackage{soul}
\usepackage{amsmath,amsfonts,mathtools,bm}
\usepackage{enumitem}
\usepackage{xcolor}
\usepackage{wrapfig}
\usepackage{dblfloatfix}
\usepackage{afterpage}
\usepackage{nowidow}
\usepackage{lineno}
\usepackage{tikz}

\newcommand{\plannabla}{{\overline\nabla}}
\newcommand{\planlap}{{\overline\Delta}}
\newcommand{\Rtwo}{{\mathbb{R}^2}}
\newcommand{\Rthree}{{\mathbb{R}^3}}
\newcommand{\Hess}{{\overline{\mathbf{H}}}}
\newcommand{\Hesst}{{{\overline{\mathbf{H}}^2}}}
\newcommand{\cHess}{{\mathbf{H}}}
\newcommand{\cHesst}{{\mathbf{H}^2}}
\newcommand{\Ric}{\operatorname{Ric}}

\DeclareMathOperator*{\argmin}{argmin}
\DeclareMathOperator*{\saddle}{saddle}

\newcommand{\tr}{\operatorname{tr}}
\newcommand{\Ho}{{H^1}}

\newcommand{\Ht}{{H^2}}

\newcommand{\Hf}{{H^4}}
\newcommand{\Hk}{{H^k}}

\newcommand{\Lt}{{L^2}}

\newcommand{\norm}[1]{{\left\lVert #1 \right\rVert}}

\newcommand{\dd}{\textrm{d}}
\newcommand{\dx}{\textrm{d}x}

\def\BibTeX{{\rm B\kern-.05em{\sc i\kern-.025em b}\kern-.08emT\kern-.1667em\lower.7ex\hbox{E}\kern-.125emX}}

\setcopyright{acmlicensed}
\acmJournal{TOG}
\acmYear{2020} \acmVolume{1} \acmNumber{1} \acmArticle{1} \acmMonth{1} \acmPrice{15.00}\acmDOI{10.1145/3377406}

\begin{document}

\title{A Smoothness Energy without Boundary Distortion for Curved Surfaces}

 \author{Oded Stein}
 \email{oded.stein@columbia.edu}
 \affiliation{\institution{Columbia University}
   \state{New York}
   \country{USA}
 }

 \author{Alec Jacobson}
 \email{alecjacobson@gmail.com}
 \affiliation{\institution{University of Toronto}
   \country{Canada}
 }

 \author{Max Wardetzky}
 \email{wardetzky@math.uni-goettingen.de}
 \affiliation{\institution{University of G\"ottingen}
   \country{Germany}
 }

 \author{Eitan Grinspun}
 \email{eitan@cs.toronto.edu}
 \affiliation{\institution{University of Toronto}
 	\country{Canada}
 }
 \affiliation{\institution{Columbia University}
   \state{New York}
   \country{USA}
 }

\renewcommand{\shortauthors}{Stein et al.}

\begin{abstract}
    Current quadratic smoothness energies for curved surfaces either
    exhibit distortions near the boundary due to zero Neumann boundary
    conditions, or they do not correctly account for intrinsic curvature,
    which leads to unnatural-looking behavior away from the boundary.
    This leads to an unfortunate trade-off: one can either have natural behavior
    in the interior, or a distortion-free result at the boundary, but not both.
    We introduce a generalized Hessian energy for curved surfaces,
    expressed in terms of the covariant one-form
    Dirichlet energy, the Gaussian curvature, and the exterior derivative.
    Energy minimizers solve the Laplace-Beltrami biharmonic equation,
    correctly accounting for intrinsic curvature, leading to
    natural-looking isolines.
    On the boundary, minimizers are as-linear-as-possible, which reduces
    the distortion of isolines at the boundary.
    We discretize the covariant one-form Dirichlet energy using
    Crouzeix-Raviart finite elements, arriving at a discrete formulation of the
    Hessian energy for applications on curved surfaces.
    We observe convergence of the discretization in our experiments.
\end{abstract}




\begin{teaserfigure}
 \includegraphics[width=\textwidth]{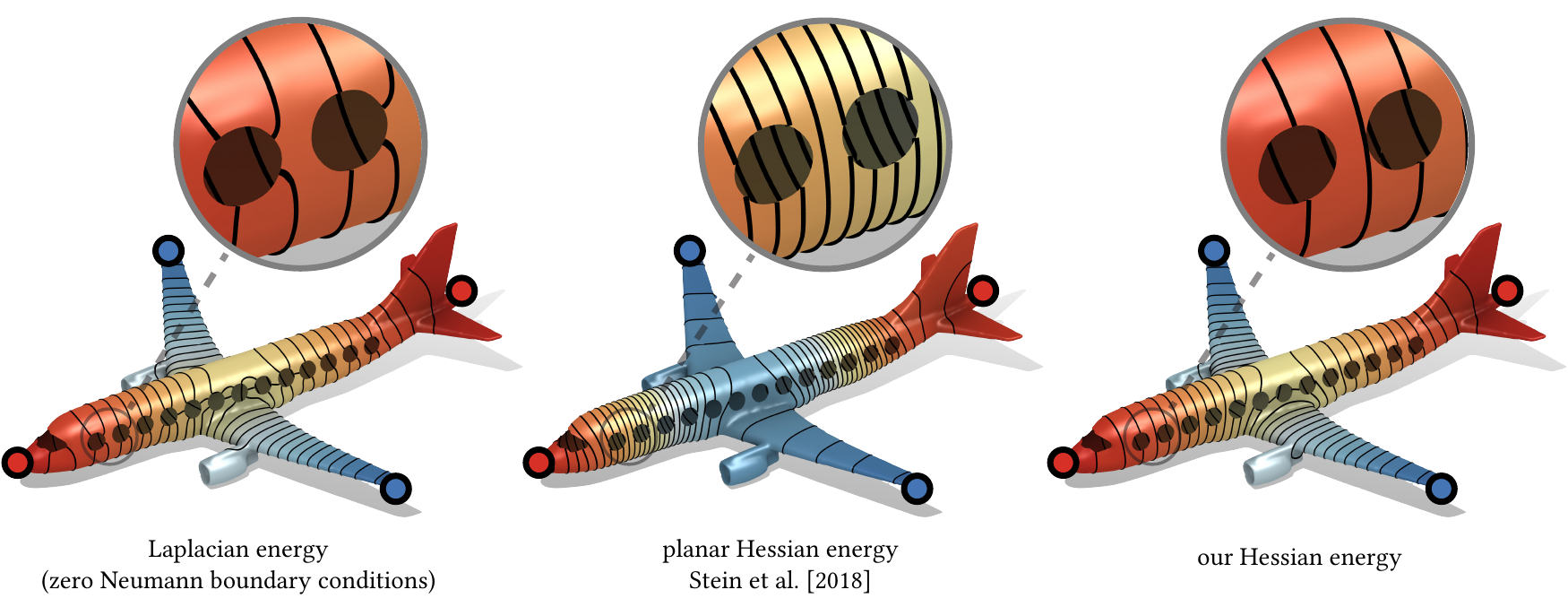}
 \caption{Solving an interpolation problem on an airplane.
 Using the Laplacian energy with zero Neumann boundary conditions
 \figloc{left} distorts the result near the windows and the cockpit of
 the plane: the isolines bend so they can be perpendicular to the boundary.
 The planar Hessian energy of \citet{Stein2018} \figloc{center} is
 unaffected by the holes, but does not account for curvature correctly,
 leading to unnatural spacing of isolines at the front and back of the
 fuselage.
 Our Hessian energy \figloc{right} produces a natural-looking result with
 more regularly spread isolines, unaffected by the boundary.
 }
 \label{fig:teaser}
\end{teaserfigure}
 
\maketitle

\section{Introduction}
\label{sec:introduction}

Smoothness energies are used as objective functions for optimization in
geometry processing.
A wide variety of applications exists:
smoothness energies can be used to smooth data on surfaces, to denoise data,
for scattered data interpolation, character animation, and much more.
We are interested in quadratic smoothness energies formulated on triangle
meshes.

	It is desirable for a smoothing energy to have minimizers with isolines
	whose spacing does not vary much across the surface---the gradient of the
	function is sufficiently constant.
	When the gradient of the function is sufficiently constant, the function
	only changes very gradually, resulting in a smooth function.
	In the same vein, a good smoothing energy should have minimizers whose
	isolines are not distorted anywhere:
	their spacing is not influenced (on the interior) by the surface's
	curvature, and they are not biased by the boundary of the surface---they
	behave locally as if the boundary were absent.
	Such behavior is relevant for applications
	where the boundary is not directly related to the actual problem that is
	being solved, e.g., when the boundary is an artificial result of faulty
	surface reconstruction resulting in a shape with many extraneous holes.
One class of energies with the desired behavior in the interior
are energies whose minimizers solve the biharmonic equation,
the prototypical elliptic equation of order four.
\cite[viii]{GazzolaGrunauSweers}.
Such energies are pertinent as smoothness energies in computer graphics
applications \cite{Jacobson2010}.

One such energy is the \emph{squared Laplacian energy}---the squared Laplacian
of a function integrated over the surface.
We henceforth refer to the energy as simply the Laplacian energy.
Its minimizers solve the biharmonic equation;
as a result, they are very smooth, and their isolines behave well on
curved surfaces, if the surfaces are closed.
The energy's most popular discretization, however, comes with zero Neumann
boundary conditions.
Thus, if a surface has boundaries, the minimizers are distorted near
the boundary (see \figref{teaser}), since at the
boundary they are \emph{as-constant-as-possible}.

The issue of boundary distortion is addressed by the \emph{Hessian energy}
of \citet{Stein2018}.
For planar domains, they provide an energy whose
minimizers solve the biharmonic equation and are \emph{as-linear-as-possible}
at the boundary.
These boundary conditions lead to decreased distortion.
The Hessian energy of \citet{Stein2018}, however, is only defined for subsets
of the plane \(\Rtwo\).
\citet{Stein2018} offer a way to compute an energy for curved
surfaces, but, as they point out, their approach does not account for the
curvature of the surface correctly.
The approach of \citet{Stein2018} does not solve the biharmonic equation
on curved surfaces;
this leads to global distortions in the isolines of the solution
(see \figref{teaser}).
\hfill\break

\subsection*{Contributions}
\paragraph{(1) Generalized Hessian energy.}
We generalize the Hessian energy to accommodate curved surfaces.
Our new Hessian energy is
\begin{equation}\label{eq:newhessianenergy}
    E(u) \coloneqq \frac{1}{2} \int_\Omega \left(\nabla \dd u \right) :
    \left(\nabla \dd u \right)
    + \kappa \left| \dd u \right|^2 \; \dx
    \;\textrm{,}
\end{equation}
where \(\nabla\) is the covariant derivative of differential forms, \(\dd\)
is the exterior derivative, \(\kappa\) is the Gaussian curvature,
and \(:\) denotes the contraction of two operators in all indices
that corresponds to {\linebreak} \(A : B = \tr(A^\intercal B)\)
(where the transpose \(\,^\intercal\) takes the metric into account).
This energy
\begin{itemize}
    \item corresponds to the Laplacian energy
    in the case of a domain without boundaries;
    \item corresponds to the Hessian energy of \citet{Stein2018}
    for surfaces in \(\Rtwo\),
    \(\frac12 \int_\Omega \lVert{\Hess_u\rVert}_F^2 \;\dx\),
    where \(\Hess_u\) is the \(2 \times 2\) Hessian matrix of \(u\), and
    \(\norm{A}_F\) is the Frobenius norm of \(A\);
    \item has the as-linear-as-possible natural boundary conditions
    of the Hessian energy of \citet{Stein2018}
    for flat domains
    in \(\Rtwo\).
    These boundary conditions lead to decreased distortion at the boundary.
\end{itemize}
\figref{teaser} shows how our Hessian energy manages to achieve the best
of both worlds.

\paragraph{(2) Discretization.}
We also introduce a discretization of this curved Hessian energy that
uses Crouzeix-Raviart finite elements ``under the hood'', but, after the energy
matrix has been assembled, relies solely on piecewise linear hat functions.
We observe convergence of the discretization for a wide variety of numerical
experiments, given certain regularity conditions,
and apply it to various smoothing and interpolation problems.
\hfill\break

\section{Related Work}
\label{sec:relatedwork}

This work extends \citet{Stein2018}.
They introduce a smoothness energy with higher-order boundary conditions
whose minimizers are biased less by the shape of the boundary than energies
using low-order boundary conditions such as zero Neumann.
Our goal is to extend their approach to curved surfaces.
Section 5.3.1 mentions that their work does not correctly account for
curved surfaces, and this shortcoming is addressed in this work.

\subsection{Smoothing energies}
	Smoothing energies are used for many applications in computer graphics,
	image processing, machine learning, and more.
	Quadratic smoothing energies are particularly interesting, since they
	are easy to work with and fast to optimize \cite{NocedalWright}.
	The Laplacian energy is used
	for surface fairing and surface editing
	\cite{Desbrun1999,Sorkine2004,Botsch2004,Crane2013uy},
	for geodesic distance computation \cite{Lipman2010},
	for creating weight functions used as coordinates in character animation
    \cite{Jacobson2011,Weber2012}, data smoothing \cite{Weinkauf2010},
	image processing \cite{Georgiev2004}, and other applications
	\cite{Jacobson2010,Sykora2014}.

	Geometric energies that share some of the properties of our Hessian energy
	have been studied in the past:
	in image processing, Hessian-like energies are popular for their
	boundary behavior, but their formulations in general do not extend to
	curved surfaces \cite{Didas2009,Lysaker2003,Lefkimmiatis2011}.
	Similar energies are also used for data processing and machine learning,
	but are not discretized for polyhedral meshes there \cite{Donoho2003,Kwang2009}.
	\citet{Wang2015,Wang2017} explicitly enforce boundary conditions on a
	\emph{discrete} quadratic fourth-order energy in order to make minimizers
	of the energy less dependent on the boundary shape, but do not
	discuss any continuous model corresponding to their method or which
	equations their minimizers satisfy.
	
	\citet{Stein2018}
	present a Hessian energy for triangle meshes, however, minimizers of their
	discretization extended to \(\Rthree\) do not fulfill the biharmonic
	equation, leading to artifacts that are discussed in detail
	in \secref{results}.
	\citet{Liu2015}
	explicitly enforce higher-order boundary conditions on a
	smoothness energy based on a fourth-order PDE.
	Their energy, however, is in general not quadratic, and the boundary
	conditions are different than the ones presented in this article, as
	they are missing the as-\allowbreak linear-as-possible property.

	A special case of a quadratic smoothness energy is the Dirichlet energy,
	which solves the harmonic equation \(-\Delta u = 0\), a simpler version of
	our biharmonic equation \(\Delta^2 u = 0\).
	The Dirichlet energy can be used, for example, to create smooth character
	deformations \cite{Joshi2007,Baran2007,Weber2007},
	and for image processing \cite{Levin2004}.
	While the Dirichlet energy has advantages, such as a discrete maximum
	principle, which is preserved in some discretizations \cite{Wardetzky2007},
	there are disadvantages due to the energy being first-order:
	because of reduced freedom around constraints, minimizers fail to be
	smooth, which can lead to artifacts when applied to shape deformation
	\cite[Fig.\ 9]{Jacobson2011}, or worse results in image processing
	\cite{Peter2016}.
	Higher-order smoothness energies, such as the ones derived from the
	biharmonic equation, are better at fitting to existing data, and tend
	to distort results less
	\cite{Georgiev2004,Jacobson2011,Jacobson2012,Weber2012}.
	Additionally, the Dirichlet energy does not admit higher-order boundary
	conditions (unlike biharmonic energies), which makes it more difficult to
	use as a smoothing energy without boundary bias.

\subsection{Generalizing the Hessian energy to curved surfaces}

A main theme in our work is the difficulty of generalizing
expressions formulated on flat domains to curved surfaces.
The presence of curvature will result in an additional term in the definition
of our energy, which is absent in the planar Hessian energy of
\citet{Stein2018}.
This mirrors many other areas of geometry where, with the introduction of
curvature, properties of flat domains
cease to apply.

One such example of curvature making calculations more elaborate is parallel
transport.
While parallel transport of vectors is trivial on flat surfaces, this is no
longer true for curved surfaces.
In the presence of curvature, the parallel transport of a vector along a
closed curve might result in a different vector than the initial one
\cite[pp.\ 156-157]{petersenriemanniangeometry}.
The difficulties that this phenomenon introduces to applications are discussed,
for example, by \citet{Polthier1998,Bergou2008,Ray2009,Crane2010}.
Our discretization method simplifies the treatment of parallel
transport by employing linear finite element basis functions that
are only supported on two adjacent triangles.
since this necessitates discontinuous basis functions, this approach is less
common.

Another instance of difficulties arising from the curved setting occurs in
the numerical analysis of finite element methods.
In order to apply standard finite element methods to curved surfaces, the
discretization has to account for the curvature of the surface.
For the case of the Poisson equation, for example,
this can be either achieved by inscribing all the vertices on the limit
surface while imposing triangle regularity conditions \cite{Dziuk1988}, or
by demanding a certain kind of convergence of the vertices as well as the
normals of the mesh \cite{Hildebrandt2006,Wardetzky2006} together with specific
triangle regularity conditions.
Similarly, in some of our own numerical experiments, we require vertex
inscription and the triangle regularity condition to achieve convergence.

\subsection{Discretization of the vector Dirichlet energy}
\label{sec:relatedworkvectordir}

An important part of the discretization of our curved Hessian energy is the
discretization of the \emph{vector Dirichlet energy}
\(\frac{1}{2} \int_\Omega \nabla \mathbf{v} : \nabla \mathbf{v} \;\dx \),
where \(\nabla\) is the \emph{covariant derivative}.
The problem of discretizing the covariant derivative for surfaces in general,
and the vector Dirichlet energy on surfaces in particular, are active areas of
research.
\citet{Knoppel2013}
provide a finite element discretization of the vector Dirichlet energy that
places the degrees of freedom on mesh vertices.
This discretization is used to design direction fields.
A different discretization, reminiscent of finite differences, can be found
in the work of \citet{Knoppel2015},
where it is used to compute stripe patterns on surfaces.
The same discretization is also used by \citet{Sharp2018}
to compute the parallel transport of vectors.
The work of \citet{Sharp2018}
also features the Weitzenb\"ock identity that we use to derive the natural
boundary conditions of our Hessian energy:
they use it to construct a Dirichlet energy on the covector bundle.
\citet{Liu2016} discretize the covariant derivative using the notion of
discrete connections.
They use it to improve the quality of the vector fields produced by
\citet{Knoppel2013}, and provide some evidence of convergence.
Other examples of discretizations of the covariant derivative include
\citet{Azencot2015}, who compute the directional derivatives of each of the
vector field's component functions,
and \citet{Corman2019}, who leverage a functional representation to compute
covariant derivatives.
\hfill\break

To simplify computation, we propose an alternative discretization of
the vector Dirichlet energy.
We use the scalar Crouzeix-Raviart finite element, the
``simplest nonconforming element for the discretization of second order
elliptic boundary-value problems'' \cite[p.\ 109]{BraessFiniteElements}.
It was first introduced by \citet{Crouzeix1973} and has become a very popular
finite element for the nonconforming discontinuous Galerkin method.
It is known to converge for the scalar Poisson equation in \(\Rtwo\).
Unlike the discretizations mentioned above, the degrees of freedom are placed
on the mesh edges.
The Crouzeix-Raviart finite element has been popular in computer graphics
applications such as the works of \citet{Bergou2006,English2008,Brandt2018};
\citet[Section 4.2]{Vaxman2016}.

Crouzeix-Raviart elements are simpler than the finite elements of
\citet{Knoppel2013}, but they come at a cost:
the basis functions are discontinuous, and the method cannot be used for
applications where the vectors have to live on vertices.
In our application, the vector-valued functions are only intermediates, so
we have more freedom in choosing their discretization, and to put vectors on
edges.

The discretization of one-forms using the Crouzeix-Raviart finite element
presented in this work is closely related to other generalizations of the
Crouzeix-Raviart element to vector- and differential-form-like quantities
such as those present in the work of \citet{Wardetzky2006},
and those discussed in the survey of \cite{Brenner2015}.
\hfill\break

\section{Smoothness energies}
\label{sec:smoothnessenergies}

A classical smoothness energy for a surface \(\Omega \subseteq \Rthree\) is the
\emph{Laplacian energy with zero Neumann boundary conditions}.
When using this method, one solves the optimization problem
\begin{equation}\begin{split}\label{eq:laplacianenergy}
	\argmin_u &\;
	\underbrace{\frac{1}{2}\int_\Omega \left| \Delta u \right|^2 \;\dx
	\quad\quad
	\frac{\partial u}{\partial\mathbf{n}} |_{\partial\Omega} = 0
	}_{E_{\Delta^2}(u)}
    \;\textrm{,}
\end{split}\end{equation}
where \(\Delta\) is the Laplace-Beltrami operator, and
\(\frac{\partial u}{\partial\mathbf{n}} |_{\partial\Omega}\) is the normal
derivative at the boundary.
\(\frac{\partial u}{\partial\mathbf{n}} |_{\partial\Omega} = 0\) is the
zero Neumann boundary condition.
In practice, when minimizing this energy by directly discretizing it and then
optimizing the resulting quadratic form, the boundary conditions manifest as
an implicit penalty on the gradient of the function at the boundary during
optimization.
We will refer to the whole optimization problem with zero Neumann
boundary conditions by \(E_{\Delta^2}\).
Minimizers of the Laplacian energy solve the biharmonic equation
\(\Delta^2 u = 0\).
This leads to natural-looking, smooth results on the interior.\footnote{Of course, simply minimizing \equref{laplacianenergy} results in the
zero function.
However, when combined with additional Dirichlet boundary conditions,
this gives a nontrivial result for the biharmonic equation
\(\Delta^2 u = 0\), and, when combined with the additional energy term
\(\int_\Omega uf \;\dx \) it gives a result for the biharmonic Poisson-type
equation \(\Delta^2 u = f\).}
The energy is easy to discretize even for meshes that are non-planar
using methods such as the mixed finite element method (FEM)
\cite{Jacobson2010}.
Using this method, the zero Neumann boundary condition does not need to
be imposed on top of the discretization, it is simply ``baked in'' by
squaring the classical cotan Laplacian.
The cotan Laplacian is also known as the Lagrangian linear FEM for the
Poisson equation
(it goes back to \citet{duffin1959} and \citet{MacNeal1949},
and its convergence for the
Poisson equation was shown by \citet{Dziuk1988}).

The minimizers of \(E_{\Delta^2}\), however, are biased by the shape of the
boundary.
Their isolines are significantly distorted near the domain boundary: they
are perpendicular to it as they have to fulfill the zero Neumann boundary
conditions (\emph{as-constant-as-possible}).
Simply removing the zero Neumann boundary conditions, and
minimizing the Laplacian energy without any boundary conditions is
not a good alternative.
Minimizations without explicit boundary conditions lead to
\emph{natural} boundary conditions.
The natural boundary conditions of the Laplacian energy are
\emph{too} permissive \cite[Fig.\ 3]{Stein2018}.
This behavior is one of the motivations for the \emph{Hessian energy} of
\citet{Stein2018}.
It is formulated as the following minimization problem.
For a surface \(U \subseteq \Rtwo\),
\begin{equation}\begin{split}\label{eq:planarhessianenergy}
	\argmin_u 
	\underbrace{
		\frac{1}{2} \int_U \Hess_u : \Hess_u \;\dx
	}_{E_{\Hesst}(u)}
\end{split}\end{equation}
where \(\Hess_u\) is the \(2 \times 2\) Hessian matrix of \(u\), and
\(\mathbf{A}:\mathbf{B} = \tr \left( \mathbf{A}^\intercal \mathbf{B} \right) \).
Minimizers of this energy solve the biharmonic equation in \(\Rtwo\).
Its natural boundary conditions lead to \emph{as-linear-as-possible}
behavior on the boundary.
This makes minimizers less biased than the zero Neumann boundary condition.

\begin{figure}
	\includegraphics[width=\linewidth]{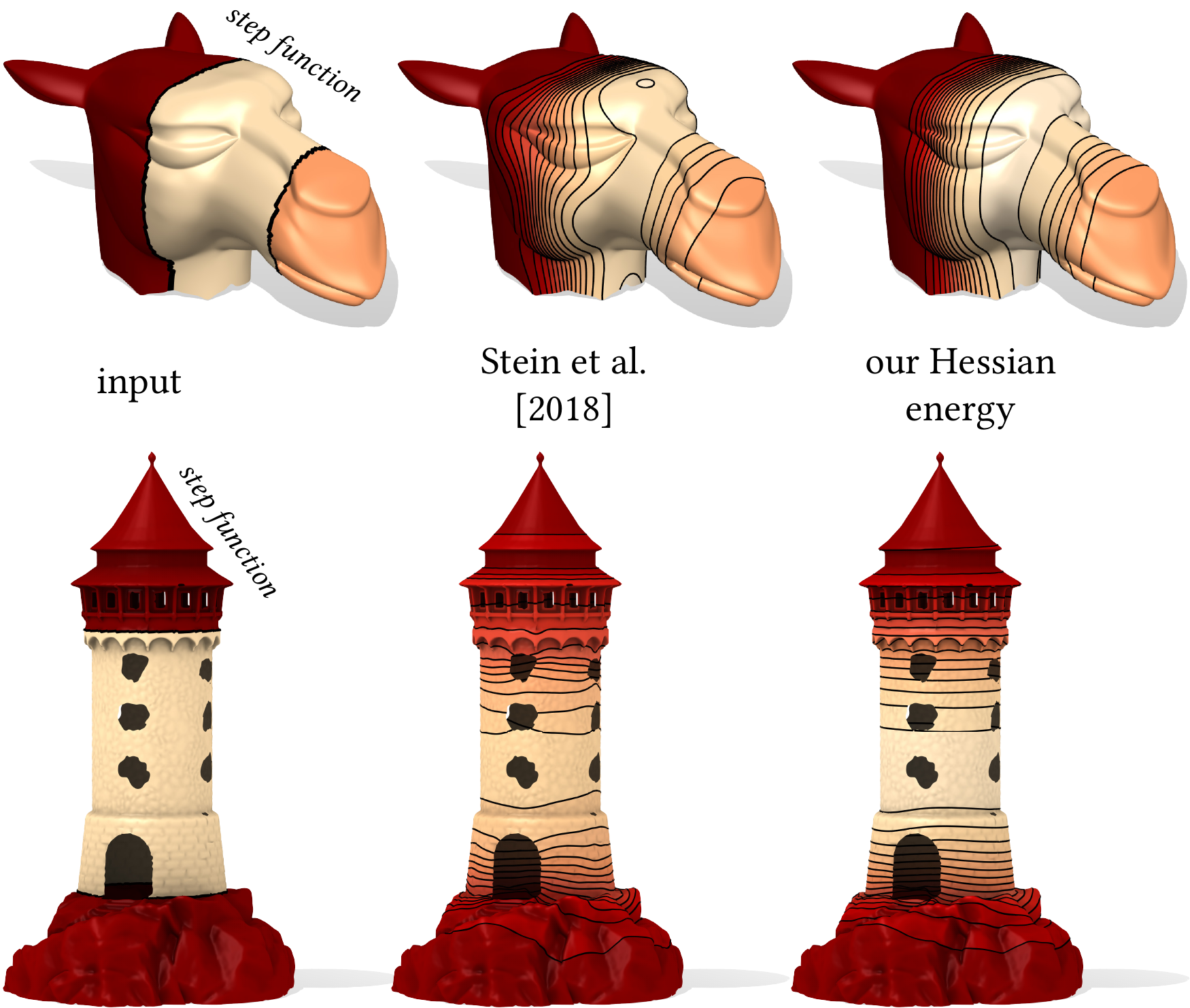}
	\caption{
		Smoothing a step function \figloc{left} on a surface
using the method of \citet{Stein2018} \figloc{middle} does not
		correctly account for the curvature of the surface, leading to crooked
		isolines.
		Our curved Hessian energy \(E\) \figloc{right} correctly accounts for
		curvature and does not suffer from such problems.
	}
	\label{fig:smoothingresult}
\end{figure}

\citet{Stein2018} demonstrate the benefits of the natural boundary conditions
of the Hessian energy with applications for curved surfaces in \(\Rthree\)
as well.
Their discretization of the planar Hessian energy for curved surfaces is
achieved by extending every operator involved in the \(\Rtwo\)
discretization to three dimensions.
This approach (the discretization, as well as the smooth formulation) does not
account for the curvature of surfaces correctly, and its minimizers do not
solve the biharmonic equation on curved surfaces
\cite[Section 5.3.1]{Stein2018}.
We refer to this generalization as the \emph{planar Hessian energy}
\(E_\Hesst\) when talking about it in the context of curved surfaces.
This planar Hessian energy is suitable for some applications, but leads to
global deviations from the natural-looking isolines produced by
\(E_{\Delta^2}(u)\) (see \figref{teaser}) or an implementation of the Hessian
energy which does account for curvature (see \figref{smoothingresult}) in
others.
\hfill\break

\section{Warm-up: the Dirichlet Energy on Curved Surfaces}
\label{sec:warmup}

As a warm-up, we consider the simple and well-known Dirichlet energy:
it is easy to generalize to curved surfaces.
We will perform the calculation for this generalization here.
The calculation is well-known, and this didactic exercise will inform our
generalization of the planar Hessian energy to curved surfaces later.

\subsection{From the energy to the PDE}

Let \(\Hk\) denote the Sobolev space of real-valued functions with \(k\) weak
derivatives in \(\Lt\).
The Dirichlet energy for domains
\(U \subseteq \mathbb{R}^2\) is defined,
for \(u \in \Ht(U)\), as\footnote{
		We choose to formulate this energy for \(u \in \Ht(U)\), although
		it is well-defined for  \(u \in \Ho(U)\), since we will continue
		our calculations with the same \(u\) right away, and we will need
		additional smoothness.
}
\begin{equation}
    E_{\plannabla^2}(u) := \frac{1}{2} \int_\Omega \plannabla u
    \cdot \plannabla u \;\dx
    \;\textrm{,}
\end{equation}
where \(\plannabla\) is the vector of partial derivatives of \(u\),
\(\plannabla u = \left( \partial_x u \;\;\; \partial_y u
\right)^\intercal \),
the normal two-dimensional gradient in \(\Rtwo\).

Minimizers of the Dirichlet energy solve the Laplace equation \cite{evans}.
Indeed, consider the variation
\begin{equation}\label{eq:adirvariation}
    u \;\rightarrow\; u + hv \quad\quad u,v \in \Ht(U)
\end{equation}
for some \(h>0\).
Since our functions are in the Sobolev function space \(\Ht\), we can differentiate them at least twice.
Plugging the variation into \(E_{\plannabla^2}(u)\), differentiating with
respect to \(h\), and then setting \(h=0\), we can see that a minimizer
\(u\) must fulfill the equation
\begin{equation*}
    \int_U \left( \partial_i u \right) \left( \partial_i v \right) \;\dx = 0
    \quad\quad \forall v \in \Ht(U)
    \;\textrm{,}
\end{equation*}
where \(\partial_*\) is a partial derivative, and summation over repeated
indices is implied.
This is a standard technique of variational calculus.
Using integration by parts (where \(\mathbf{n}\) is the boundary normal)
\begin{equation}\begin{split}\label{eq:fullgradderivt}
    0 &= \int_U \left( \partial_i u \right) \left( \partial_i v \right) \;\dx
    \\
    &= \int_{\partial U} \left( \partial_i u \right) v \; \mathbf{n}_i \;\dx
    - \int_U \left( \partial_i \partial_i u \right) v \;\dx
    \;\textrm{.}
\end{split}\end{equation}
Here a boundary term appeared as a result of integration by parts. 
The second term of the second line corresponds to the standard
two-dimensional planar Laplacian
\(\planlap = \plannabla \cdot \plannabla\),
and so we conclude that minimizers of the energy \(E_{\plannabla^2}(u)\)
fulfill the two-dimensional planar Laplace equation \(-\planlap u = 0\).
The additional boundary term, the first term of the second line in
\equref{fullgradderivt}, determines the \emph{natural boundary conditions} of
the Dirichlet energy.
They are called natural boundary conditions because they naturally emerge
from solving the variational problem over the set of all functions, without
explicitly enforcing additional boundary conditions. 
In this case we can see that the natural boundary conditions are zero-Neumann
boundary conditions:
\begin{equation}\label{eq:flatdirichletnaturalbc}
    \partial_i u \; \mathbf{n}_i
    = \plannabla u \cdot \mathbf{n} = 0 \quad\quad \textrm{on }
    \partial U
    \;\textrm{.}
\end{equation}

\subsection{From the PDE to a new energy}
\label{sec:dirichletpdetoenergy}

We now generalize the Dirichlet energy to curved surfaces.
This means we are looking for an energy whose minimizers solve a curved
version of the Laplace equation, and fulfill a curved version of the natural
boundary conditions \equref{flatdirichletnaturalbc}.
	While we were able to write the calculations in terms of coordinates
	in the flat setting, this is much harder to do in the curved setting.
	This is why we perform calculations in the curved setting in a
	coordinate-free fashion. 

The curved analog of the planar Laplace equation is
\(\Delta u = 0\), where \(\Delta\) is the \emph{Laplace-Beltrami operator}
\cite[Chapter 3]{JostRiemannian}.
It holds for a function \(u \in \Ht(\Omega)\) 
(where \(\Omega\) is a compact surface immersed in \(\Rthree\).) that
\begin{equation}
    \Delta u = \delta \dd u
    \;\textrm{,}
\end{equation}
where \(\dd\) is the exterior derivative and \(\delta\) is the codifferential,
the (formal) dual of the exterior derivative under integration by parts.
For planar surfaces, the Laplace-Beltrami operator \(\Delta\)
corresponds to \(-\planlap\).

We start with an integral formulation of the Laplace equation, and then
use integration by parts.
For all \(v \in \Ht(\Omega)\) it must hold that
\begin{equation*}\begin{split}
    0 &= \int_\Omega \left(\Delta u\right) v \;\dx
    = \int_\Omega \left(\delta\dd u\right) v \;\dx \\
    &=
    - \int_{\partial\Omega} \langle \dd u, \mathbf{n} \rangle \; v \;\dx
    + \int_\Omega \left(\dd u\right) \cdot \left(\dd v\right) \;\dx
    \;\textrm{,}
\end{split}\end{equation*}
where the natural (metric-independent) pairing of one-forms and vectors is
indicated using the angle bracket, and \(\cdot\) is the dot product of
one-forms.

Using the definition of the gradient \(\nabla\) on curved surfaces,
\(\nabla u \cdot \mathbf{w} \coloneqq \langle \dd u, \mathbf{w} \rangle\) for
a vector \(\mathbf{w}\) (where \(\cdot\) is the dot product of vectors
and the angle bracket \(\langle \cdot , \cdot \rangle\) denotes the pairing
of a one-form with a vector)
\cite[(3.1.16)]{JostRiemannian}, we can write
\begin{equation}\begin{split}\label{eq:weakcurveddirichlet}
    0 &=
    - \int_{\partial\Omega}  \nabla u \cdot \mathbf{n} \; v \;\dx
    + \int_\Omega \nabla u \cdot \nabla v \;\dx
    \;\textrm{.}
\end{split}\end{equation}

Walking back through the variation from \equref{adirvariation},
this now motivates the definition of a curved Dirichlet energy
\begin{equation}
    E_{\nabla^2}(u) \coloneqq \frac{1}{2} \int_\Omega \nabla u \cdot \nabla u \;\dx 
    \;\textrm{.}
\end{equation}
We have shown that minimizers of this energy solve the curved Laplace
equation, and by the boundary term in \equref{weakcurveddirichlet} it is also
clear that minimizers fulfill a curved zero Neumann boundary condition:
\begin{equation}
    \nabla u \cdot \mathbf{n} = 0 \quad\quad \textrm{on } \partial U
    \;\textrm{.}
\end{equation}

Thus we have successfully generalized the Dirichlet energy to curved
surfaces.
Even though we went through the work of using differential geometric operators,
we ended up with something quite similar to what we started with, but with
\(\plannabla\) replaced by \(\nabla\).
For more complicated energies this will no longer be the case.
\hfill\break

\section{The Hessian Energy on Curved Surfaces}
\label{sec:thehessianoncurvedsurfaces}

We now seek to derive a smooth Hessian energy on surfaces that
generalizes the Hessian energy in \(\Rtwo\),
while ensuring that minimizers of the energy solve the biharmonic equation.
This will follow the approach we used in \secref{warmup} to generalize the
planar Dirichlet energy to curved surfaces.

\subsection{From the energy to the PDE}

For the planar Hessian energy \(E_\Hesst\) it is a straightforward calculation
to prove that minimizers fulfill the biharmonic equation.
This calculation is mentioned, for example, in \citet[Section 4]{Stein2018},
and we will repeat it here for convenience.
Our setting is a compact planar domain \(U \subseteq \Rtwo\).
The linear equation fulfilled by minimizers of \equref{planarhessianenergy}
derived with standard variational calculus is:
find \(u \in \Hf(U)\) such that
\begin{equation}\label{eq:variationalplanarhessian}
   \int_U (\partial_i \partial_j u)  (\partial_i \partial_j v) \;\dx = 0
   \quad\quad \forall v \in \Hf(U)
    \;\textrm{,}
\end{equation}
where, as before, \(\partial_*\) is a partial derivative, and summation
over repeated indices is implied.
Using integration by parts (where \(\mathbf{n}\) is the boundary normal)
we know that
\begin{equation}\begin{split}\label{eq:flatbiharmderivation}
   0 &= \int_U (\partial_i \partial_j u)  (\partial_i \partial_j v) \;\dx \\
   &= \int_{\partial U} (\partial_i \partial_j u) (\partial_j v)
   \mathbf{n}_i \; \dx
    - \int_U (\partial_i \partial_i \partial_j u)  (\partial_j v) \;\dx \\
    &= \int_{\partial U} (\partial_i \partial_j u) (\partial_j v)
   \mathbf{n}_i - (\partial_i \partial_i \partial_j u) v \mathbf{n}_j \; \dx \\
   &\quad\quad + \int_U (\partial_j \partial_i \partial_i \partial_j u) v \;\dx
   \;\textrm{.}
\end{split}\end{equation}

Since all partial derivatives commute in the plane, in the very last term we
can write \(\partial_j \partial_i \partial_i \partial_j u
= \partial_i \partial_i \partial_j \partial_j u = \planlap^2 u\).
As a result, we can conclude that minimizers of the Hessian energy satisfy
the biharmonic equation with some additional boundary terms.
This commutation will not be that easy for curved surfaces.

After some rearranging, these boundary terms can be seen to imply the
natural boundary conditions
\begin{equation}\begin{split}\label{eq:naturalconditionsplane}
    \mathbf{n}^\intercal \Hess_u \mathbf{n} &= 0
    \quad\quad \textrm{on } \partial U \\
    \plannabla\planlap u \cdot \mathbf{n} + \plannabla
    \left(\mathbf{t}^\intercal \Hess_u \mathbf{n} \right) \cdot \mathbf{t}
    &= 0
    \quad\quad \textrm{on } \partial U
    \;\textrm{,}
\end{split}\end{equation}
where \(\mathbf{n}\) is the normal vector at the boundary, and \(\mathbf{t}\)
is the tangential vector of the (oriented) boundary.
A derivation of \equref{naturalconditionsplane} can be found
in the work of \citet[Section 4.3]{Stein2018}.
\hfill\break

\textbf{A naive approach to a Hessian energy for curved surfaces.}
Since our goal is to generalize the Hessian energy for surfaces, it seems 
natural to simply replace the planar Hessian \(\Hess_u\) with an analog for
curved surfaces, and minimize this generalization of the Hessian energy.
Unfortunately, this will not work: the resulting minimizers of such an
energy will not solve the biharmonic equation.

Consider a compact surface \(\Omega\) immersed in \(\Rthree\).
We define the Hessian of a function on a curved surface
\cite[p.\ 54]{leeriemannianmanifolds}
\begin{equation}
    \cHess_u := \nabla \dd u
    \;\textrm{,}
\end{equation}
where \(\nabla\) applied to one-forms is the covariant derivative of
differential forms and \(\dd\) is the exterior derivative.
It might seem reasonable to define a generalized Hessian energy as
\begin{equation}\label{eq:naivehessianenergy}
    E_\cHesst(u) := \frac{1}{2} \int_\Omega \cHess_u : \cHess_u \;\dx
    \;\textrm{,}
\end{equation}
where \(:\) now denotes the contraction of all indices.
The associated variational equation at a stationary point is
\begin{equation*}\int_\Omega \left(\nabla \dd u\right) : \left(\nabla \dd v\right) \;\dx = 0
   \quad\quad \forall v \in \Hf(\Omega)
   \textrm{.}
\end{equation*}
We can already see that we will not be able to repeat our approach from
\equref{flatbiharmderivation}:
there is no way to easily commute \(\nabla\) and \(\dd\),
as it was possible in the flat setting with coordinate-wise calculation,
and thus we can't perform the same simple calculation to show that minimizers of
\(E_\cHesst\) solve the biharmonic equation.

\subsection{From the PDE to a new energy}

Instead, echoing \secref{dirichletpdetoenergy}, we derive an
energy whose minimizers fulfill the boundary conditions
\equref{naturalconditionsplane} and also solve the biharmonic equation.
We start with the integrated biharmonic equation using the Hodge Laplacian
operator \(\Delta = \dd \delta + \delta \dd \) for forms on surfaces, which
degenerates to the Laplace-Beltrami operator \(\delta \dd\) for zero-forms
(scalar functions), and which corresponds to the standard Laplacian for
functions in the plane.
It holds that
\begin{equation}\begin{split}\label{eq:curvedhessianderiv1}
    0 &= \int_\Omega \left(\Delta\Delta u\right) v \;\dx
    = \int_\Omega \left(\delta\dd\delta\dd u\right) v \;\dx \\
    &=
    - \int_{\partial\Omega} \langle \dd\delta\dd u, \mathbf{n}\rangle
    \; v \; \dx
    + \int_\Omega \left(\dd\delta\dd u\right) \cdot \left( \dd v \right) \; \dx
    \;\textrm{,}
\end{split}\end{equation}
where \(\mathbf{n}\) is the boundary normal vector, and we used the fact
that the exterior derivative \(\dd\) is dual to the codifferential \(\delta\).

\begin{figure}
 \includegraphics[width=\linewidth]{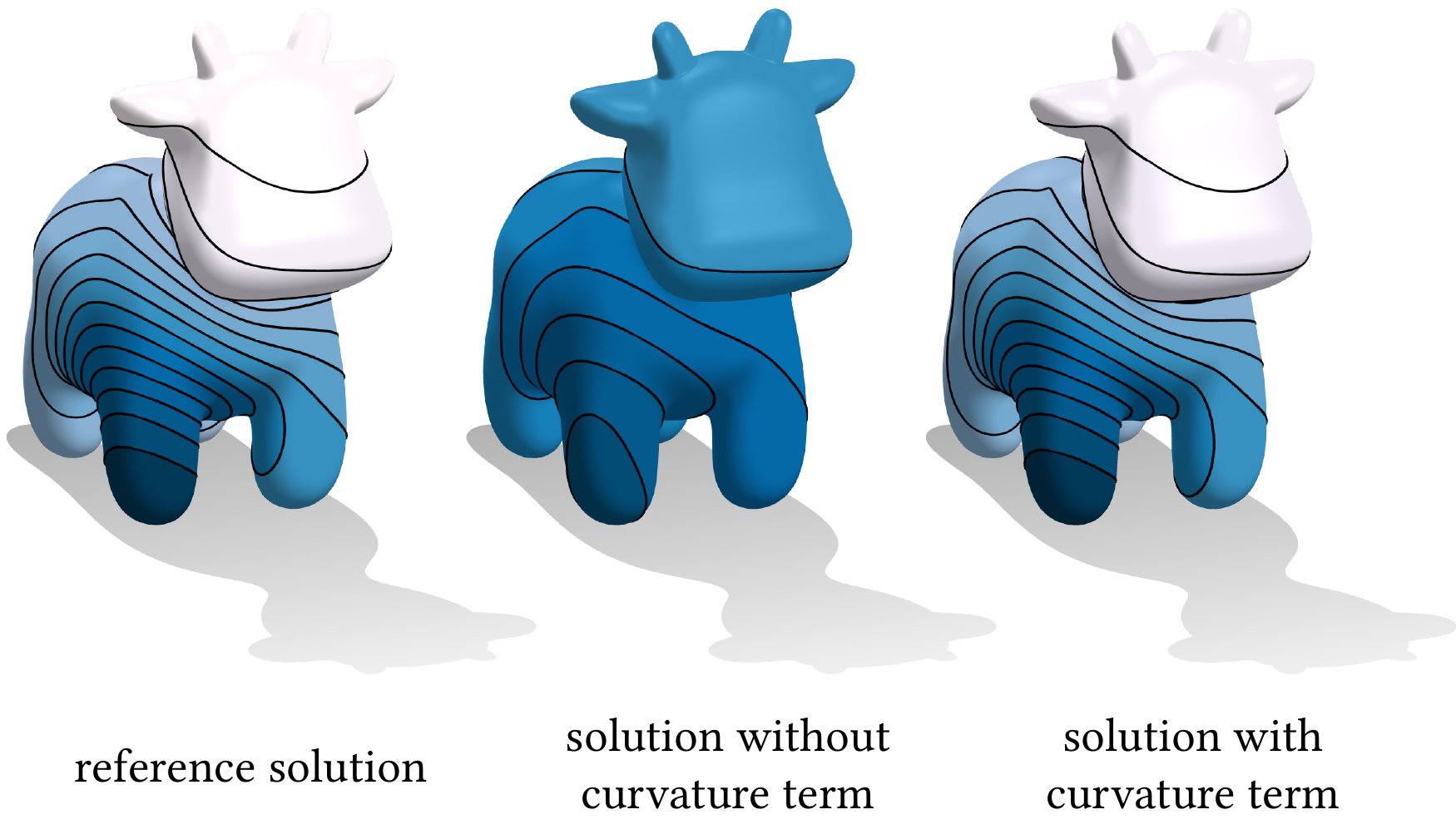}
 \caption{We solve the Poisson-like problem \(\Delta^2 u = f\) using the
 Hessian energy with \figloc{right, \(E\)} and without
 \figloc{center, \(E_\cHesst\)} curvature term.
 The solution for \(E_{\Delta^2}\) is provided as a reference solution
 \figloc{left}.
 We see that the solution for \(E\) corresponds to the reference solution,
 since its minimizers solve the biharmonic equation, while
 the solution for \(E_\cHesst\) does not.}
 \label{fig:withandwithoutcurvaturecorrection}
\end{figure}

Now we utilize the Weitzenb\"ock identity.
It relates the Hodge-Laplacian \(\Delta = \dd \delta + \delta \dd \) and the
Bochner Laplacian \(\Delta_B = \nabla^* \nabla\), where \(\nabla^*\) is the
(formal) dual covariant derivative.
	The formal dual is defined via integration by parts on a closed manifold
	\(M\),
	\(\int_M X : \nabla \omega \;\dx = \int_M \nabla^* X \cdot \omega \;\dx\).
It holds that
\begin{equation}
    \Delta = \nabla^* \nabla + \Ric
    \;\textrm{,}
\end{equation}
where \(\Ric\) is the Ricci curvature tensor
\cite[Chapter 7]{petersenriemanniangeometry}.
This formula dates back to \citet{Bochner1946} and \citet{Weitzenbock1885}.
It is used, together with the fact that \(\dd^2=0\), to continue our
calculation from \equref{curvedhessianderiv1}.
\begin{equation}\begin{split}\label{eq:curvedhessianderiv2}
    \int_\Omega \left(\dd\delta\dd u\right)& \cdot \left( \dd v \right) \; \dx
    = \int_\Omega \left((\dd\delta + \delta\dd)\dd u\right) \cdot
    \left( \dd v \right) \; \dx \\
    &= \int_\Omega \left(\nabla^*\nabla\dd u\right) \cdot
    \left( \dd v \right) + \Ric(\dd u, \dd v) \; \dx \\
    &= 
    - \int_{\partial\Omega} \mathbf{n}^i \left(\nabla\dd u\right)_{ij} \cdot
    \left( \dd v \right)_j \; \dx \\
    &\quad\quad + \int_\Omega \left(\nabla\dd u\right) :
    \left(\nabla\dd v \right) + \Ric(\dd u, \dd v) \; \dx
    \;\textrm{,}
\end{split}\end{equation}
where indices have been added to make clear which contraction happens in which
index.

The term involving the Ricci curvature tensor \(\Ric\) can be further simplified.
For the case of two-dimensional manifolds we know that we can write
the Ricci curvature tensor as simply
\begin{equation}\label{eq:simplifyricci}
    \Ric = \kappa g
    \;\textrm{,}
\end{equation}
where \(\kappa\) is the Gaussian curvature, \ie,
half the scalar curvature
\cite[pp.\ 38-41]{petersenriemanniangeometry}.

Putting \equref{curvedhessianderiv1}, \equref{curvedhessianderiv2}, and
\equref{simplifyricci} together then gives
\begin{equation}\begin{split}\label{eq:finalcurvedvariation}
    0 &= 
    - \int_{\partial\Omega} \langle \dd\delta\dd u, \mathbf{n}\rangle
    \; v + \mathbf{n}^i \left(\nabla\dd u\right)_{ij} \cdot
    \left( \dd v \right)_j \;\dx \\
    &\quad\quad + \int_\Omega \left(\nabla\dd u\right) :
    \left(\nabla\dd v \right) + \kappa \, \dd u \cdot \dd v \; \dx
    \;\textrm{.}
\end{split}\end{equation}
This is, in the case of a planar surface (for which it holds \(\kappa=0\)),
exactly the term from our earlier calculation with the planar Hessian energy
from \equref{flatbiharmderivation}.
Here we also see why minimizers of the naive Hessian energy \(E_\cHesst\)
do not solve the biharmonic equation on curved surfaces:
the energy \(E_\cHesst\) lacks the curvature correction term
involving \(\kappa\)
(see \figref{withandwithoutcurvaturecorrection}).

The result from \equref{finalcurvedvariation} motivates the definition of the
following curved Hessian energy:
\begin{equation}\label{eq:repnewhessianenergy}
\boxed{
    E(u) := \frac{1}{2} \int_\Omega \left(\nabla \dd u \right) :
    \left(\nabla \dd u \right)
    + \kappa \left| \dd u \right|^2 \; \dx
}
    \;\textrm{.}
\end{equation}
Minimizers of the energy \(E\) solve the biharmonic equation on a
surface in \(\Rthree\), unlike minimizers of \(E_\cHesst\).
\hfill\break

It remains to check what the natural boundary conditions of \(E\) are.
We can find them by checking which biharmonic functions \(u\) fulfill the
boundary terms
\begin{equation*}\begin{split}
    0 &= \int_{\partial\Omega} \langle \dd\delta\dd u, \mathbf{n}\rangle
    \; v + \mathbf{n}^i \left(\nabla\dd u\right)_{ij} \cdot
    \left( \dd v \right)_j \;\dx
    \quad \forall v \in \Hf(\Omega)
    \;\textrm{.}
\end{split}\end{equation*}
We use the same strategy as \citet[Section 4.3]{Stein2018}:
testing with specific subsets of all valid test functions.
These subsets are purpose-built to
expose the natural boundary conditions of the energy.
First, we test with all functions \(v\) that vanish on the boundary, and
thus only have nonzero differential in the normal direction
(\(v=0, \; \langle \dd v, \mathbf{w} \rangle
= g \mathbf{n} \cdot \mathbf{w}\) for some smooth \(g\)).
It follows that
\begin{equation}\label{eq:naturalcond1}
    \mathbf{n}^i \left(\nabla\dd u \right)_{ij} \mathbf{n}^j = 0
    \quad\textrm{on } \partial\Omega
    \;\textrm{,}
\end{equation}
i.e., the (curved) Hessian of the solution is linear across the boundary;
	the second derivative of the function across the boundary is zero.
This mirrors the ``as-linear-as-possible'' condition of
\citet[(17)]{Stein2018}.

Using the same strategy of testing the expression with a specific subset of
functions to expose boundary behavior, if we plug in
all functions that have zero differential in the normal
direction at the boundary (\(\langle \dd v, \mathbf{n} \rangle = 0\)), we get
\begin{equation}\label{eq:naturalcond2}
    \langle \dd\delta\dd u, \mathbf{n} \rangle
    + \delta_{\partial\Omega,j} \imath_{\partial\Omega}
    \left( \mathbf{n}^i \left(\nabla\dd u\right)_{ij}
    \right) = 0
    \quad\textrm{on } \partial\Omega
    \;\textrm{,}
\end{equation}
where \(\imath_{\partial\Omega}\) is the natural projection of one-forms on the
surface to one forms on the boundary, and the subscript on the codifferential
implies that this is the codifferential of the boundary manifold in the
index \(j\).
This mirrors the condition from \citet[(18)]{Stein2018}.
In fact, the two natural boundary conditions \equref{naturalcond1} and
\equref{naturalcond2} of the Hessian energy are exactly the ones of the planar
Hessian energy if the domain is a planar surface.
\hfill\break

\begin{figure}
 \includegraphics[width=\linewidth]{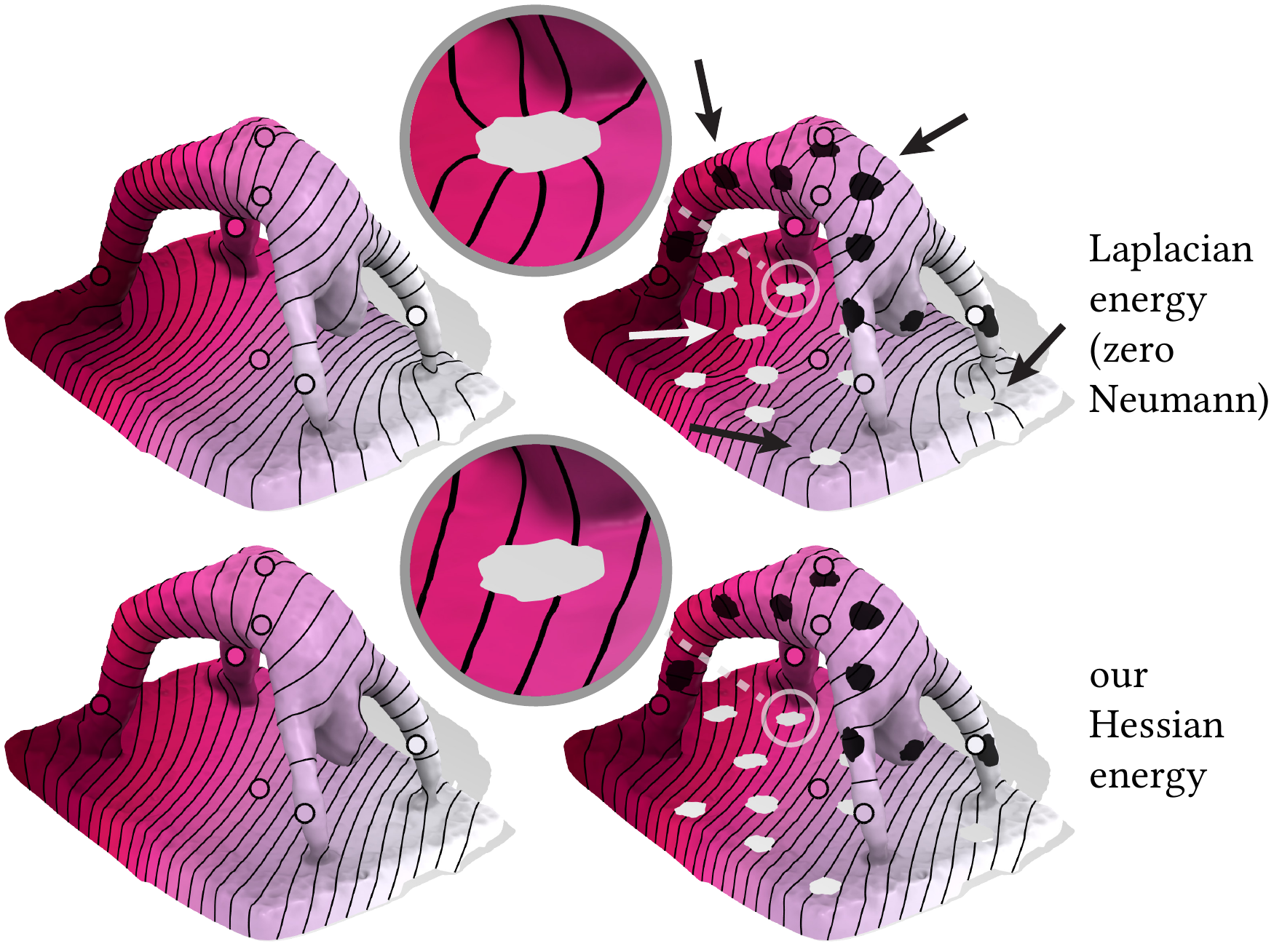}
 \caption{Using the Laplacian energy \(E_{\Delta^2}\) \figloc{top} for
 scattered data interpolation gives a result that is influenced by the
 boundary:
 adding holes makes the isolines near them bend towards the holes.
 Our Hessian energy \(E\) \figloc{bottom} is less distorted at the holes and
 produces a very similar result without and with holes.}
 \label{fig:scatteredcomparisonwithholed}
\end{figure}

\subsubsection*{The Hessian energy natural boundary conditions}
Like the natural boundary conditions of \(E_\Hesst\) from
\citet[Section 4.3]{Stein2018}, the natural boundary conditions
\equref{naturalcond1} and \equref{naturalcond2} of the Hessian energy
\(E\) guarantee that its minimizers
\begin{itemize}
    \item continue linearly across the boundary in the normal direction
    (\(\mathbf{n}^i \left(\nabla\dd u \right)_{ij} \mathbf{n}^j = 0\)),
    and
    \item have limited variation along the boundary
    \quad\quad\quad\quad\quad\quad\quad\quad\(\;\)
    (\(\langle \dd\delta\dd u, \mathbf{n} \rangle
    + \delta_{\partial\Omega,j} \imath_{\partial\Omega}
    \left( \mathbf{n}^i \left(\nabla\dd u\right)_{ij}
    \right) = 0\)),
\end{itemize}
as discussed by \citet[Section 4.3]{Stein2018}.
Both boundary conditions are fulfilled by minimizers of \(E\)
in the absence of explicitly enforced boundary conditions.

On planar surfaces, these boundary conditions mean that the null space of
the energy contains all linear functions, in contrast to the Laplacian
energy with zero Neumann boundary conditions \(E_{\Delta^2}\), whose null space 
only contains constant functions.
On closed surfaces, the null space of \(E\) and \(E_{\Delta^2}\) is the same:
all constant functions.

The natural boundary conditions of the Hessian energy have a physical
interpretation.
Consider a deforming flat thin plate where displacement is modeled by the
function \(u\).
The plate is not clamped or supported at the boundary in any way:
it is a \emph{free plate}.
Then the conditions \equref{naturalcond2} are the boundary conditions
fulfilled by \(u\) \cite[pp.\ 206-207]{CourantHilbert}.
These boundary conditions go back at least as far as
\citet[p.\ 355]{Rayleigh1894}.

Its natural boundary conditions make the Hessian energy a good choice for
ignoring the boundary as much as possible, while maintaining biharmonic
behavior everywhere away from the boundary (see
\figref{scatteredcomparisonwithholed} where they are contrasted with zero
Neumann boundary conditions).
\hfill\break

\begin{figure}
\includegraphics[width=\linewidth]{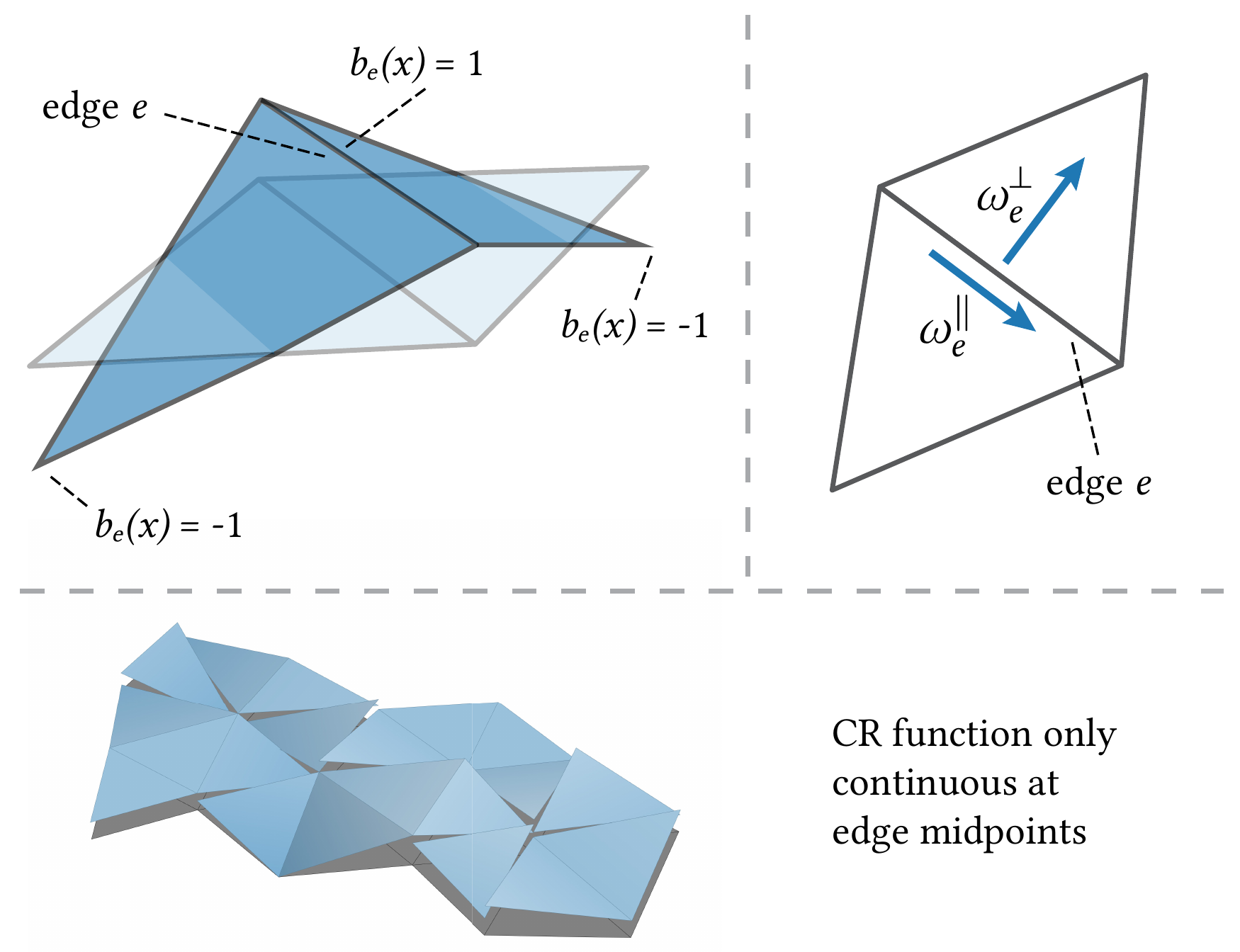}
	\caption{A scalar Crouzeix-Raviart basis function for the edge \(e\)
		\figloc{top left}.\\
		The parallel and perpendicular one-forms for the edge \(e\), represented
		by their dual vectors \figloc{top right}.\\
		Crouzeix-Raviart functions and their sums are, in general,
		discontinuous.
		Continuity is only guaranteed at edge midpoints
		\figloc{bottom}.
	}
	\label{fig:crouzeixraviartbasisfct}
\end{figure}

\section{Discretization}
\label{sec:discretization}

We offer a discretization for the curved Hessian energy \(E\) derived in
\secref{thehessianoncurvedsurfaces}.
The approach presented here is a simple method using only linear finite
elements, intended to make the Hessian energy easily accessible.
There are, however, other conceivable ways to discretize this energy,
such as, for example, higher-order conforming finite elements
\cite[II.5]{BraessFiniteElements}.

\subsection{Computing the Hessian energy}
\label{sec:computingthehessian}

Discretizing the Hessian energy \(E\) \equref{repnewhessianenergy} as written 
would require us to discretize functions that can be differentiated twice.
To avoid this complication we use the mixed finite element method
\cite{BoffiFEM} by introducing an intermediate variable \(w = \dd u\) and
formulate the problem of minimizing \(E\) as
\begin{equation}
    \label{eq:aconstrainedproblem}
    \argmin_u\;
    \frac{1}{2} \int_\Omega \left(\nabla w \right) :
    \left(\nabla w \right)
    + \kappa \left| w \right|^2 \; \dx,
    \quad w = \dd u
    \;\textrm{.}
\end{equation}
Using Lagrange multipliers to enforce the constraint \(w = \dd u\), we can
write the optimization problem as the saddle problem
(where our goal is finding a stationary point)
\begin{equation}\begin{split}
    \label{eq:lagrangifiedproblem}
    \saddle_{u,w,\lambda}\;
    \frac{1}{2} \int_\Omega &\left(\nabla w \right) :
    \left(\nabla w \right)
    + \kappa \left| w \right|^2 \; \dx \\
    &- \int_\Omega \lambda \cdot
    \left( w - \dd u \right) \; \dx
    \;\textrm{.}
\end{split}\end{equation}

We discretize the space of scalar functions (containing \(u\)) using standard
continuous, piecewise linear functions which are a very commonly used finite
element.
Definitions are found, for example, in \citet[II.5]{BraessFiniteElements}.
The basis of this discrete space consists of
the \(\varphi_i, \; i=1,\dots n\),
sometimes called ``hat functions'' (see inset).
\setlength{\columnsep}{5pt}
\setlength{\intextsep}{0pt}
\begin{wrapfigure}[7]{r}{95pt}
	\includegraphics{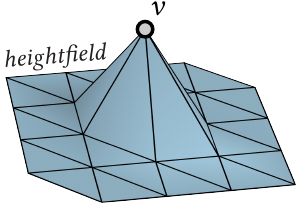}
\end{wrapfigure}
We write \(u = \sum_i u_i \varphi_i \), and we have the vector
\(\mathbf{u} = \begin{pmatrix}u_1, \dots, u_n\end{pmatrix}^\intercal\).

The space of one-forms (containing \(w\)) is discretized using Crouzeix-Raviart
one-forms (CROFs), which are described in \secref{crof}.
The basis of this discrete space are the functions \(\eta_i, \;i=1,\dots m\).
We write \(w = \sum_i w_i \eta_i\), and we have the vector
\(\mathbf{w} = \begin{pmatrix}w_1, \dots, w_m\end{pmatrix}^\intercal\).
\hfill\break

\begin{figure}
 \includegraphics[width=\linewidth]{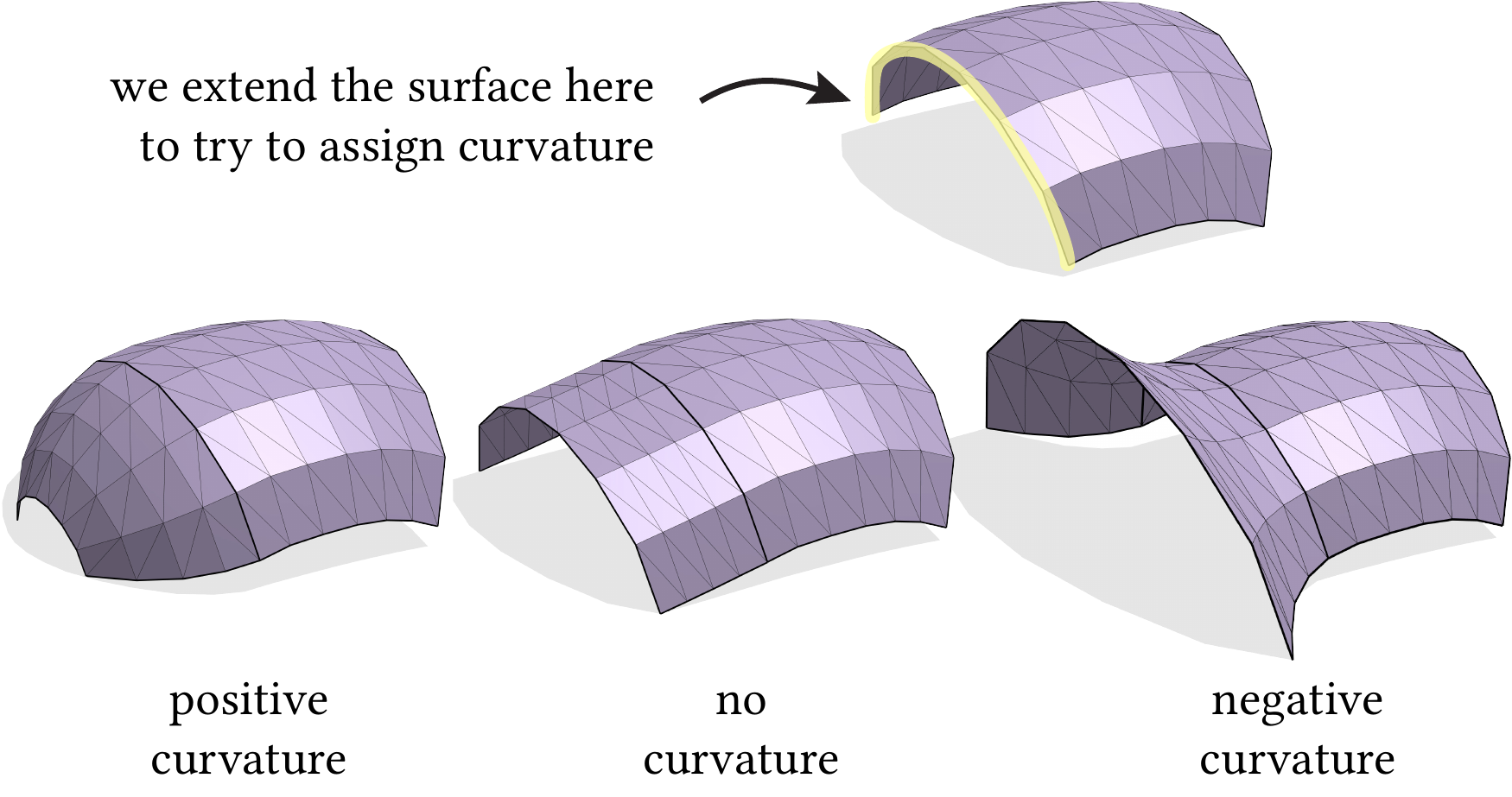}
 \caption{For the boundary of a continuous, piecewise linear surface
 \figloc{top} there is no way to uniquely assign curvature at the boundary.
 The surface can be extended in many different ways that yield different
 curvatures at the boundary, examples leading to positive
 \figloc{bottom left}, no \figloc{bottom center}, and negative
 \figloc{bottom right} curvature are shown.}
 \label{fig:curvatureatboundary}
\end{figure}

Using these discretizations we can construct the one-form Dirichlet matrix
\begin{equation*}
    L_{ij} = \int_\Omega \left(\nabla \eta_i\right) :
    \left(\nabla \eta_j\right)
    \;\dx
    \;\textrm{,}
\end{equation*}
the differential matrix
\begin{equation*}
    D_{ij} = \int_\Omega \eta_i \cdot \dd \varphi_j \;\dx
    \;\textrm{,}
\end{equation*}
the mass matrix
\begin{equation*}
    M_{ij} = \int_\Omega \eta_i \cdot \eta_j \;\dx
    \;\textrm{,}
\end{equation*}
and the curvature matrix
\begin{equation*}
    K_{ij} = \int_\Omega \kappa \eta_i \cdot \eta_j \;\dx
    \;\textrm{.}
\end{equation*}
The matrix entries are provided in \appref{implementation}.

Using these matrices, we write the discrete version of
\equref{lagrangifiedproblem} as seeking a critical point of the expression
\begin{equation*}
\frac{1}{2} \mathbf{w}^\intercal \left(L + K\right) \mathbf{w}
    - \bm{\lambda}^\intercal \left(M\mathbf{w} - D\mathbf{u}\right)
    \;\textrm{,}
\end{equation*}
for \(\mathbf{u} \in \mathbb{R}^n\),
\(\mathbf{w}, \bm{\lambda} \in \mathbb{R}^m\).
Differentiating with respect to \(\bm{\lambda}\) gives
\(M\mathbf{w} = D\mathbf{u}\).
As \(M\) is invertible, we get the system
\begin{equation}\boxed{
	\label{eq:simplifieddiscreteenergy}
    \argmin_{\mathbf{u}} \;
    \mathbf{u}^\intercal
    D^\intercal M^{-1} (L+K) M^{-1} D
    \mathbf{u}
}
\;\textrm{.}
\end{equation}
This optimization problem can now be solved with a variety of constraints,
or mixed with other energy terms, depending on the application.

\subsection{Crouzeix-Raviart One-Forms}
\label{sec:crof}

While there are multiple approaches to discretizing tangent one-forms for
triangle meshes, we choose to base
our approach on Crouzeix-Raviart finite elements
(see \secref{relatedwork} for a discussion).
The advantage of this approach is its simplicity.
Crouzeix-Raviart basis functions are only ever nonzero on two adjacent
triangles, so every basis function lives on an intrinsically flat
domain: the two triangles can be unfolded without distortion.
This means that our discretization will account for curvature
correctly in the end, without having to explicitly address issues like
parallel transport during construction.

\subsubsection{Introduction to Crouzeix-Raviart}

The scalar Crouzeix-{\linebreak}Raviart basis function for the edge \(e_{ij}\)
is defined to be \(1\) on the edge itself, \(-1\) on the two vertices \(k,l\)
opposite the edge, and linear on the two triangles \(T_{ijk}, T_{jil}\)
\cite[p.\ 109]{BraessFiniteElements}.
For boundary edges, only one triangle needs to be considered.
As a result, it is \(0\) on the midpoints of the edges \(e_{jk}, e_{ki},
e_{il}, e_{lj}\) (see \figref{crouzeixraviartbasisfct}, left).
The scalar Crouzeix-Raviart element is not continuous, except at the midpoints
of edges.
This makes it a non-conforming element, and if it is used in a Galerkin
method, one speaks of the discontinuous Galerkin method.
Despite being nonconforming, it is known to converge for certain problems,
most notably the Poisson equation in \(\Rtwo\).
\cite[III, Theorem 1.5]{BraessFiniteElements}.

\begin{figure}[b]
	\includegraphics[width=\linewidth]{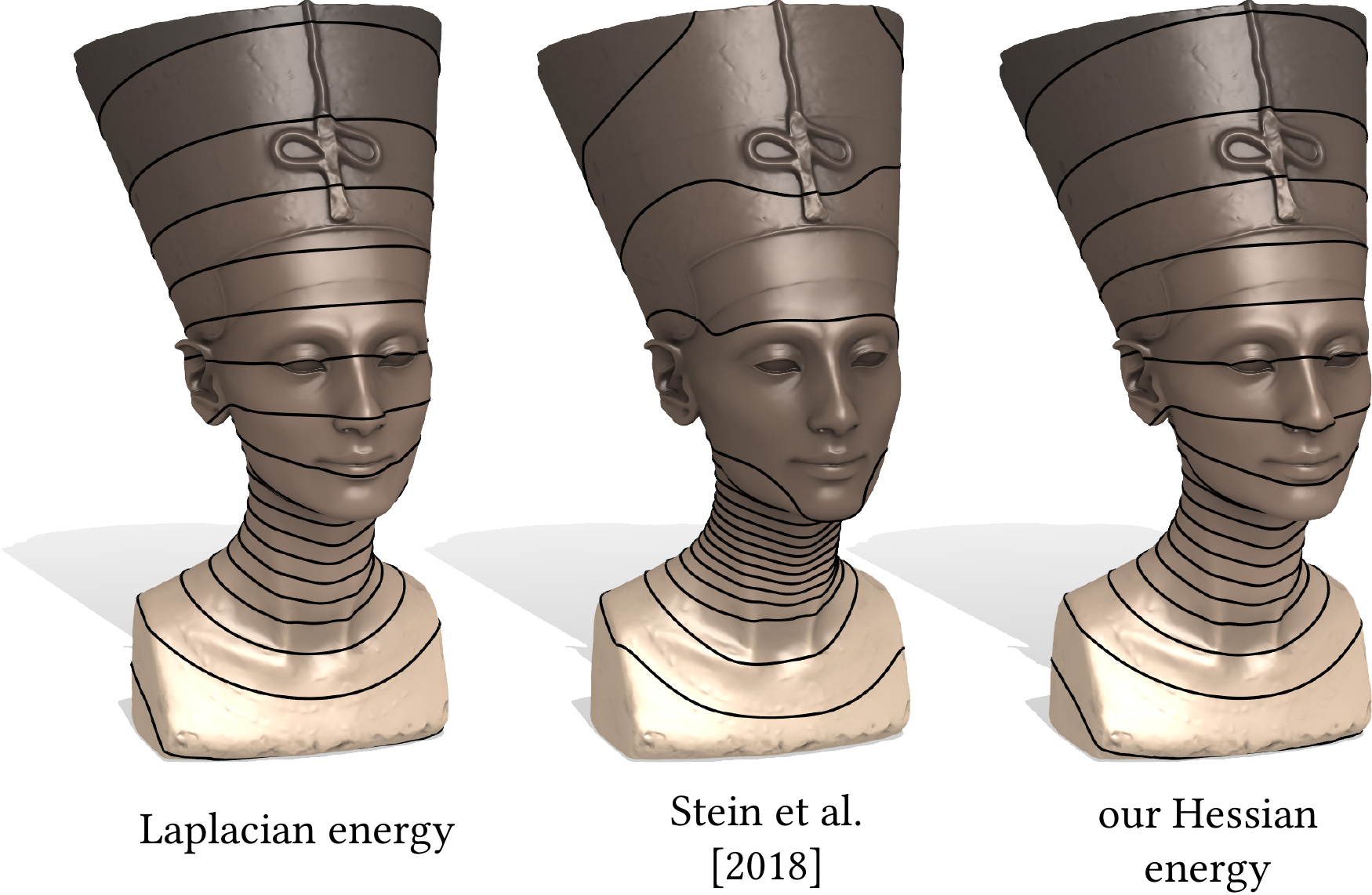}
	\caption{The first nonzero eigenvector of the Laplacian energy
		\(E_{\Delta^2}\) \figloc{left}, the Hessian of \citet{Stein2018}
		\figloc{center}, and
		the curved Hessian energy \(E\) \figloc{right}.
		The eigenvectors of \(E_{\Delta^2}\) and \(E\) look similar, since they both
		discretize the biharmonic energy.
		The method of \citet{Stein2018} visibly disagrees.}
	\label{fig:eigenvectors}
\end{figure}

\subsubsection{One-Forms}

The scalar Crouzeix-Raviart element can be used to define a finite element
space for one-forms.
At the midpoint of every edge \(e\) of a flat triangle pair, the space of
one-forms is spanned by the two forms \(\omega^\parallel_e, \omega^\perp_e\),
such that
\begin{equation}\begin{split}
    \langle \omega^\parallel_e, \mathbf{t}_e \rangle = 1,
    &\quad
    \langle \omega^\parallel_e, \mathbf{n}_e \rangle = 0
    \\
    \langle \omega^\perp_e, \mathbf{t}_e \rangle = 0,
    &\quad
    \langle \omega^\perp_e, \mathbf{n}_e \rangle = 1
    \;\textrm{,}
\end{split}\end{equation}
where \(\mathbf{n}_e\) is the (oriented) perpendicular vector of the edge
\(e\) in each triangle, \(\mathbf{t}_e\) is the (oriented) tangent of
the edge \(e\),
and the angle bracket \(\langle \cdot ,\cdot \rangle\) denotes the pairing
of a form with a vector.
See \figref{crouzeixraviartbasisfct} (right) for an illustration.
The definition of \(\omega^\perp_e\) depends on which triangle one is in,
but only in an \emph{extrinsic} way:
in the intrinsic geometry of the triangle pair, the edge is completely flat,
thus the two covectors \(\omega^\perp_e\) defined in each triangle are the
same covector intrinsically.
Because of this, both \(\omega^\parallel_e\) and \(\omega^\perp_e\)
can be easily extended to the triangles adjacent to \(e\):
since the triangle pair (or the one triangle) is intrinsically flat,
parallel transport along the triangles is trivial, and we can easily extend
the definition of \(\omega^\perp_e\) and \(\omega^\parallel_e\) to the
interior of the triangles adjacent to \(e\).

If \(b_e\) is the Crouzeix-Raviart basis function for the edge \(e\), then we
define its two CROF basis function as
\begin{equation}\begin{split}
    b_e^\parallel &\coloneqq \omega^\parallel_e b_e \\
    b_e^\perp &\coloneqq \omega^\perp_e b_e
    \;\textrm{.}
\end{split}\end{equation}

Defined this way, CROFs have the correct notion of parallel
transport, without having to explicitly account for it.
Consider a path \(\gamma\) through all edge midpoints of edges
emanating from a vertex \(v\) in a counterclockwise direction (see inset).
We start with a single tangent vector on the midpoint of one edge,
corresponding to a combination of two basis functions, and see what
angle we pick up when going around the vertex \(v\) using our basis functions.
\setlength{\columnsep}{5pt}\setlength{\intextsep}{0pt}\begin{wrapfigure}[7]{r}{88pt}\includegraphics{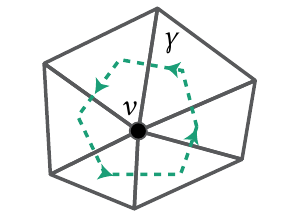}\end{wrapfigure}We now go along the path \(\gamma\), moving from
edge to edge by choosing successive basis functions so that the sum of the
basis functions from two adjacent edges is constant on the shared triangle.
Doing that corresponds exactly to parallel transport on a cone manifold:
the tangential part of the vector at each edge does not change extrinsically
at all when crossing the edge along \(\gamma\).
The perpendicular basis function jumps extrinsically: the angle between
normal vectors on each side of the edge is \(\pi\) minus the dihedral angle of
the edge.
At the end of our journey along \(\gamma\), when we are back at our original
edge, our starting vector picked up angle defect
corresponding to the discrete curvature of the mesh.
The CROF basis functions have accounted for the discrete curvature of the
mesh in the sense of curvature on cone manifolds \cite{Wardetzky2006}
without having to explicitly account for parallel transport during the
construction of the basis functions.

Since every basis function is only supported on at most two triangles, the
matrices \(L, M, D, K\) will be sparse.
The matrix \(M\) is diagonal, which makes it easy to invert.
The matrix entries can be found in \appref{implementation}.
\hfill\break

\begin{figure}[b]
	\includegraphics[width=\linewidth]{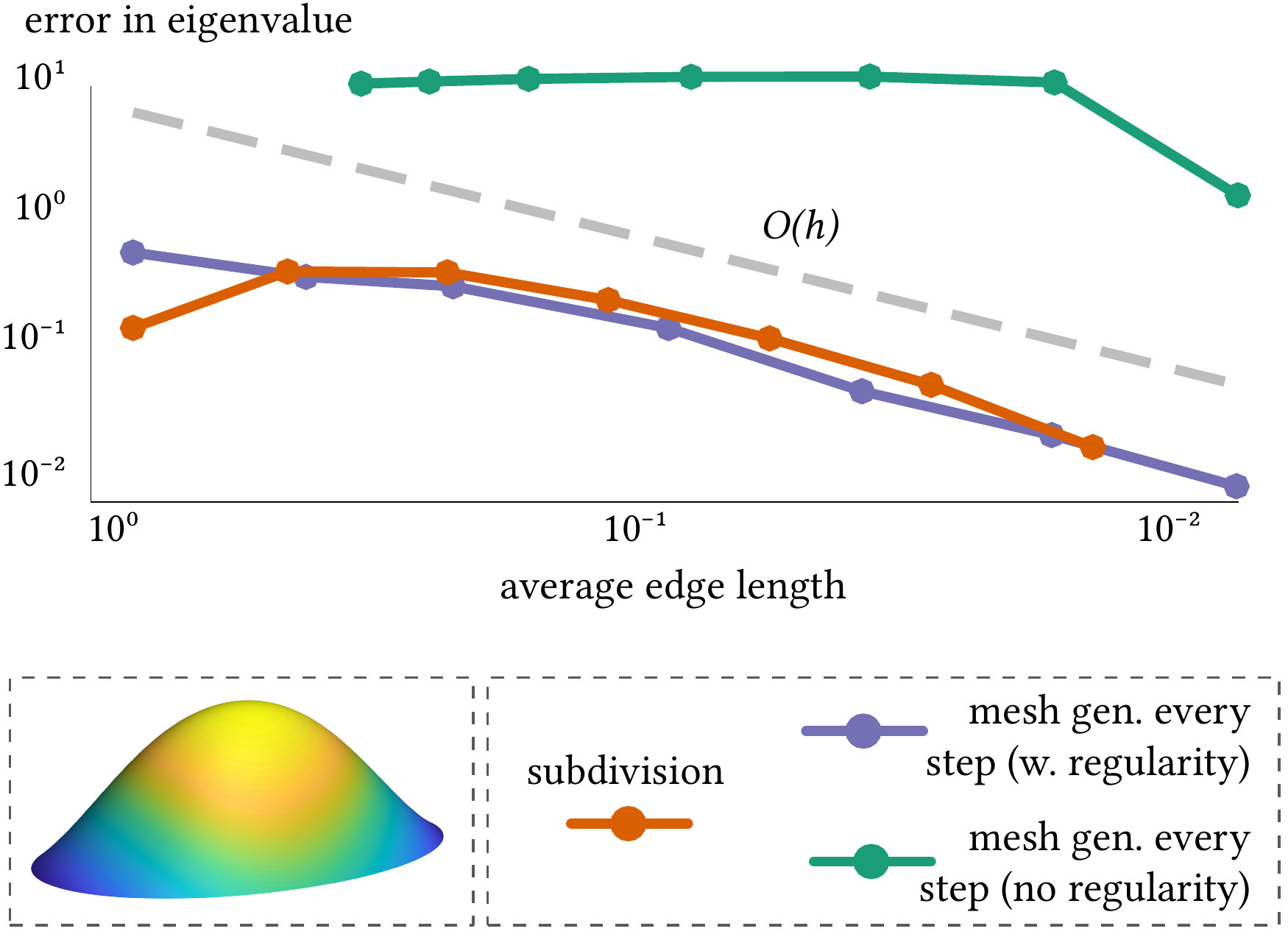}
	\caption{Computing the fourth eigenvalue of the Hessian energy \(E\) on
		an ellipse that was distorted in the third dimension
		\figloc{bottom left}.
		Both refinement through Loop subdivision and projection to a given smooth
		surface, as well as generating a planar mesh of the desired resolution 
		with regular triangles at every step and then projecting to a given smooth
		surface show convergence to the highest resolution.
		For simple mesh generation without triangle regularity no convergence is
		observed.
	}
	\label{fig:arbitraryrefinements}
\end{figure}

\begin{figure*}[b]
	\quad\\\quad\\
	\includegraphics[width=\linewidth]{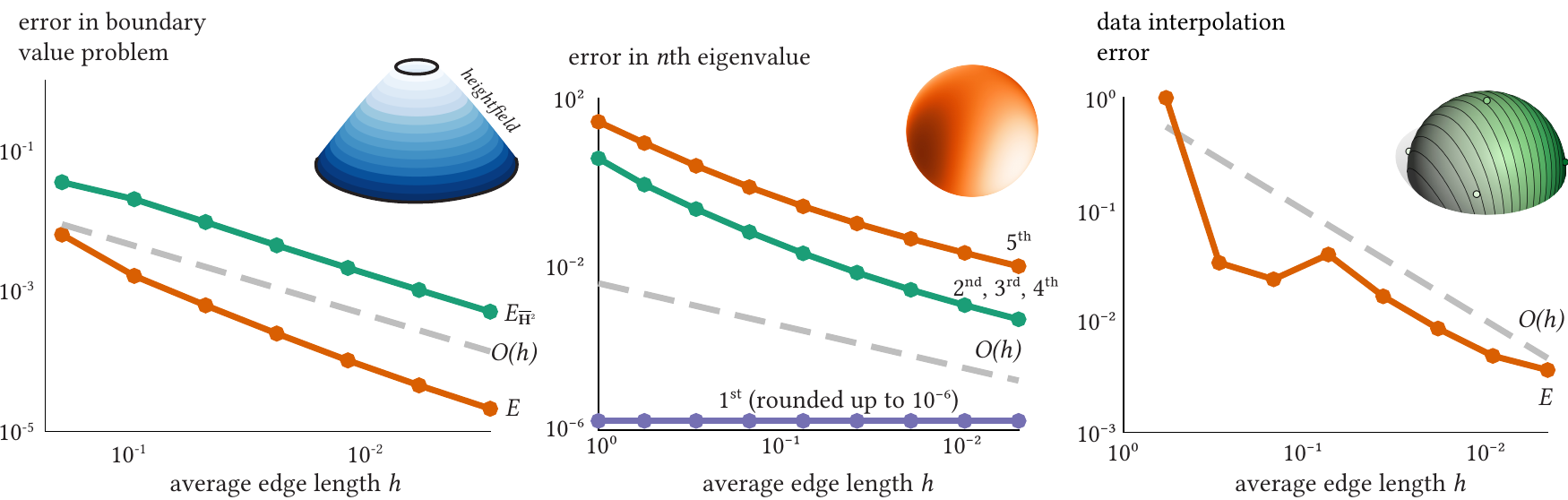}
	\caption{Convergence plots for three different problems,
		all errors are \(\Lt\) errors.
		Boundary value problem with known exact solution on a
		flat annulus mesh
		refined by loop subdivision with fixed smooth boundary; both
		our Hessian \(E\) and the planar Hessian \(E_\Hesst\) of 
		\cite{Stein2018}
		are shown
		(even though, for planar domains, the smooth curved and planar
		Hessian energies coincide, the different discretizations result
		in a different error) \figloc{left}.
		Error in calculating the lowest eigenvalues
		of the operator associated with \(E\) on the sphere with icosahedral 
		meshing, with vertices of the mesh inscribed in the smooth limit sphere
		\figloc{center}.
		Solving an interpolation problem and computing the error with respect
		to the highest-resolution solution, refined by loop subdivision with
		fixed \(z\)-coordinate at the boundary \figloc{right}.
	}
	\label{fig:convergenceplots}
\end{figure*}

\subsubsection{The curvature term}
\label{sec:curvatureterm}

Special care needs to be applied when computing the matrix \(K\).
The Gaussian curvature \(\kappa\) of an intrinsically flat pair of triangles
would appear, at first, to be \(0\).
But actually, the Gaussian curvature of a polyhedron is entirely concentrated
on its vertices (and is zero anywhere else).
The integrated Gaussian curvature at a vertex is also known as the
angle defect
\begin{equation}
    \kappa_v \coloneqq 2\pi - \sum_{f \in N(v)} \theta_v^f
    \;\textrm{,}
\end{equation}
where the sum is over all faces \(f\) in the set of faces
containing the vertex \(v\), and \(\theta_v^f\) is the angle at vertex \(v\)
in face \(f\) \cite{Grinspun2006}.
The idea of angle defects is very old:
it goes back all the way to at least Descartes c.\ 1630, who showed that
the sum of all angle defects of a polyhedron with spherical topology is
\(4 \pi\) \cite{Federico1982}.

We thus interpret the Gaussian curvature of the polyhedron as a collection of
delta functions at every vertex, i.e.\ 
\begin{equation}
    \kappa \coloneqq \sum_v \kappa_v \delta_v
    \;\textrm{,}
\end{equation}
where \(\delta_v\) is the Dirac delta.
This means that the integral of \(\kappa g\), where \(\kappa\) is the
Gaussian curvature and \(g\) is any continuous function over the triangle 
\(T_{ijk}\) with vertices \(i,j,k\), can be written as:
\begin{equation}
    \int_\Omega \kappa g \;\dx = \sum_v \kappa_v g(v)
    \;\textrm{.}
\end{equation}
If the function \(g\) itself is only continuous in each triangle, then we need
to distribute the contribution of each triangle accordingly.
Let \(s_{v,f}>0\) for each vertex \(v\) and each face \(f\)
in the neighborhood of \(v\) be coefficients that average the contribution
of each face at a vertex, \ie, the sum of the \(s_{v,f}\)
over all faces \(f\) in the neighborhood of \(v\) is one.
Then
\begin{equation*}
    \int_\Omega \kappa g \;\dx = \sum_v \kappa_v
    \sum_{f \in N(v)} s_{v,f} \;g_f(v)
    \;\textrm{,}
\end{equation*}
where \(g_f\) is the function \(g\) in the triangle \(f\)
and \(N(v)\) is the set of all faces in the neighborhood of \(v\).
We choose to average by tip angle,
which corresponds to an integral along a small circle around the vertex.
We did not explore other reasonable choices, such as averaging by face area.
This formula
is used to compute the entries of \(K\), they are given in
\appref{implementation}.

One remaining issue with the angle defect as Gaussian curvature is that the
angle defect is not defined at boundary vertices.
The problem stems from the fact that the notion of curvature at the boundary
of meshes (continuous, piecewise linear surfaces) is not in and of itself
meaningful:
by choosing to extend the surface in different ways at the boundary we can
achieve any arbitrary Gaussian curvature, as can be seen in
\figref{curvatureatboundary}.
We choose to set the angle defect to \(0\) for all boundary
vertices, thereby choosing the \emph{most developable} (intrinsically linear)
extension of all possible extensions.
This fits in with our \emph{as-linear-as-possible} boundary conditions,
but differs from some conventions of angle defect at the boundary, which
define it as the sum of tip angles subtracted from \(\pi\) (which is a
discretization of geodesic curvature).
\hfill\break

\subsection{Observed Numerical Convergence}
\label{sec:performance}

Using our CROF discretization of the Hessian energy to solve a variety of
problems, we observe convergence on the order of the average edge length
\(h\) (\figref{convergenceplots}).
As can be seen in \figref{arbitraryrefinements}, a successful strategy
for obtaining convergence is making sure that the vertices are contained in
a smooth surface, and then either refining the mesh through Loop subdivision
\cite{LoopSubdivision} with a fixed smooth boundary, or generating meshes
that fulfill the triangle regularity condition:
the ratio of circumcircle to incircle of each triangle (the triangle
regularity) is bounded from above and below independent of refinement level.
This condition is standard for finite elements
\cite[Definition 5.1 (uniform triangulation)]{BraessFiniteElements}.
The order of convergence and the triangle regularity condition correspond
to the discretization of the Laplacian energy
with zero Neumann boundary conditions, \(E_{\Delta^2}\), with mixed FEM
in the flat setting \cite{Scholz1978,Jacobson2010}.
However, we do not have a proof of convergence for our method to confirm
this convergence rate.

\begin{figure}
	\includegraphics[width=\linewidth]{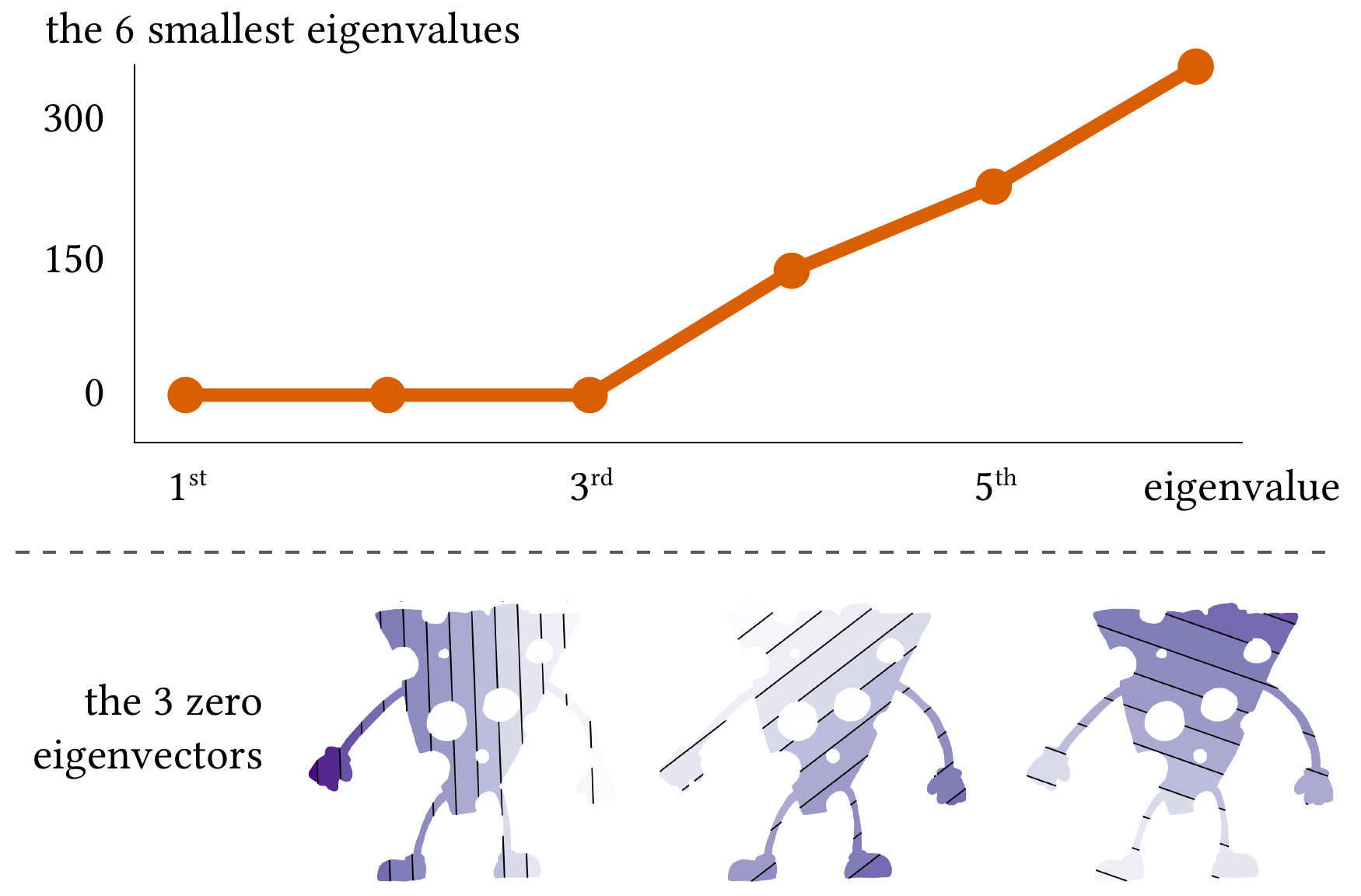}
	\caption{The six lowest eigenvalues of the Hessian energy discretized with
		CROF on the cheeseman \figloc{top}.
		As expected, there are only three zero eigenvalues.
		The three lowest eigenvectors \figloc{bottom} are the linear functions, which
		corresponds to the smooth Hessian energy.
	}
	\label{fig:lowesteigenvectors}
\end{figure}

Our method correctly reproduces the first eigenvector of the Laplacian energy
on closed surfaces in the experiment proposed by
\citet[Section 5.3.1]{Stein2018} on a refined mesh (\figref{eigenvectors}).
As mentioned in \citet[Section 4.5]{Stein2018}, discretizations can sometimes
exhibit spurious modes in the kernel of the energy, which lead to wrong
solutions.
We have not proved that this does not happen for our CROF
discretization of the Hessian energy, but we have not observed it in our experiments
(see \figref{lowesteigenvectors} for the cheeseman example domain mentioned in
\citet[p.\ 7]{Stein2018}).

Further experiments can be found in \appref{additionalexperiments}:
Figures \ref{fig:manyconvergenceplots} and \ref{fig:forwardproblems} feature
additional convergence experiments confirming the order of convergence,
\figref{meshingdependence} examines the dependence of the result on the
mesh further, and \figref{deccomparisons} compares our implementation of the
Hessian energy with other Hessian energies in the flat case.
\hfill\break

\section{Applications}
\label{sec:results}

We implement the optimization of \equref{simplifieddiscreteenergy}
by constructing a sparse matrix in \texttt{C++} using Eigen \cite{eigenweb},
and then manipulating and optimizing it in MATLAB \cite{MATLAB} with
\texttt{mex}.
For linear equality constraints, we use the optimizer of
\citet[\texttt{min\_quad\_with\_fixed}]{libigl} via the library of
\citet{gptoolbox}.
Using this approach, complicated constraints are also possible, such as
linear and quadratic inequality constraints for more complicated
applications.
Since the Hessian energy is a quadratic energy, optimizers using the
interior point method (such as the solver of \citet{mosek}) are appropriate.

\begin{figure}
	\includegraphics[width=\linewidth]{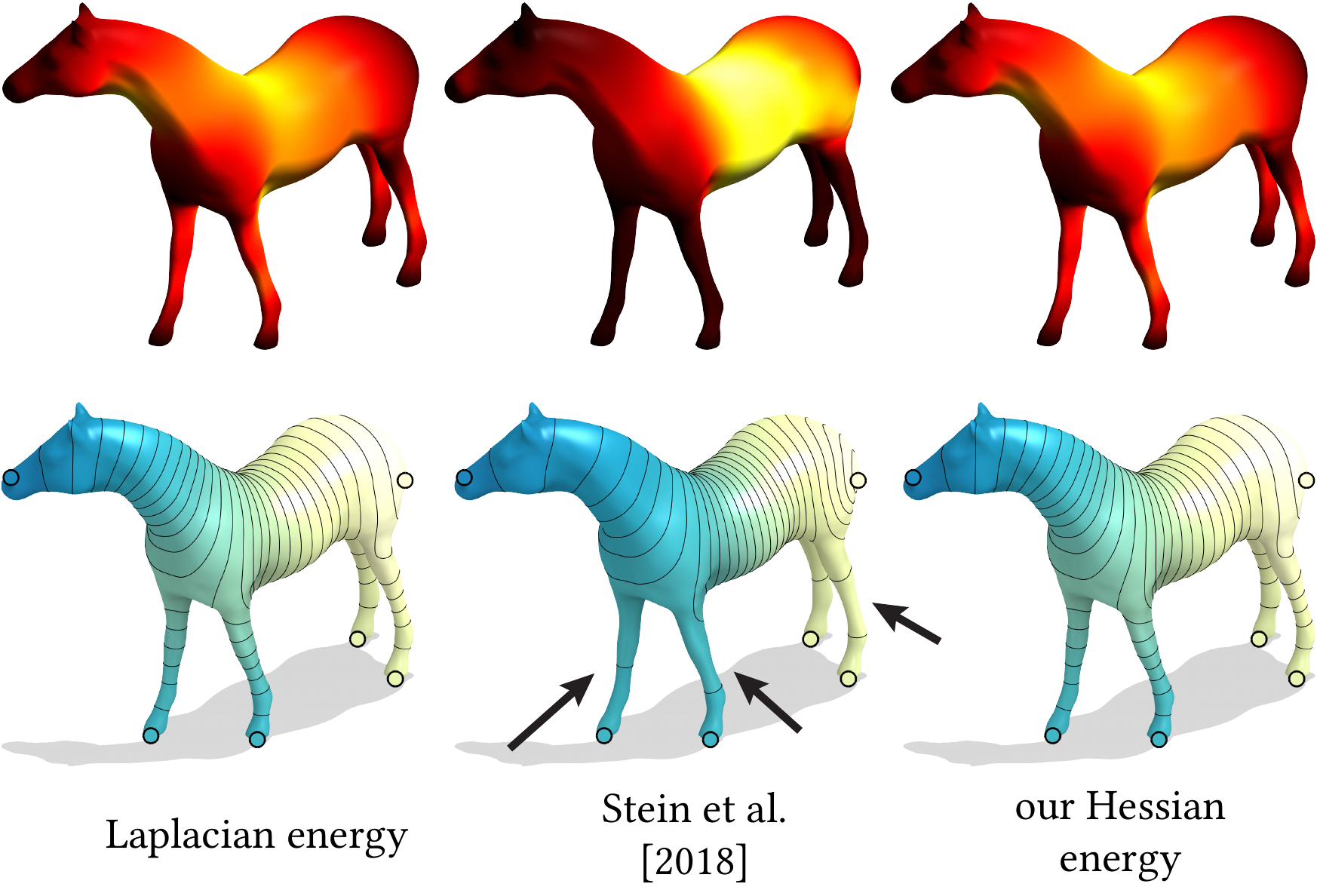}
	\caption{Scattered data interpolation problem solved on a closed surface
		\figloc{bottom row} and the gradient of the solution \figloc{top row}.
		\(E_{\Delta^2}\) \figloc{left} provides a satisfying result---isolines
		are relatively evenly spaced, and the gradient is uniform.
		\citet{Stein2018} \figloc{center} has large variation in
		isoline distance (see arrows), and the gradient of the solution is less
		uniform.
		\(E\) \figloc{right} replicates the behavior of \(E_{\Delta^2}\).}
	\label{fig:scatteredcomparisonclosedsurface}
\end{figure}

\begin{figure*}
	\includegraphics[width=\textwidth]{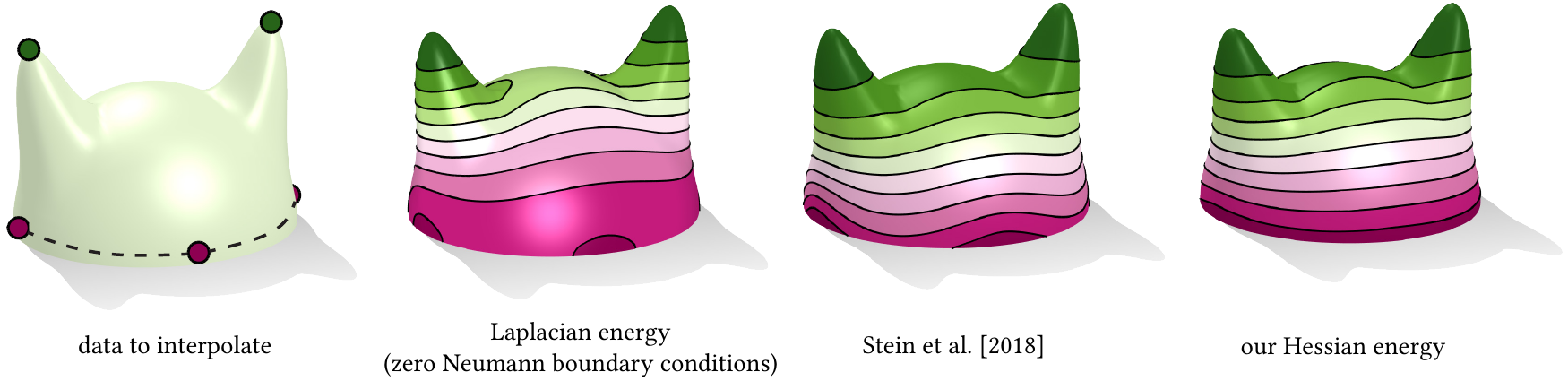}
	\caption{Solving an interpolation problem on a viking helmet.
		Our goal here is to preserve the dashed line (which is almost a geodesic)
		connecting three data points of the same value \figloc{far left}.
		Using \(E_{\Delta^2}\) distorts the line near the boundary, since the
		zero Neumann boundary conditions make the isolines
		perpendicular to the boundary \figloc{center left}.
		Using the planar Hessian of \citet{Stein2018} still leads to some
		distortion due to not accounting for the surface's curvature
		\figloc{center right}.
		Our Hessian energy \(E\) correctly accounts for the curvature of the
		surface and does not suffer from bias at the boundary, interpolating the
		dashed line as desired \figloc{far right}.}
	\label{fig:vikinghelmet}
\end{figure*}

\subsection{Scattered data interpolation}
\label{sec:scattereddatainterpolation}

Like any smoothness energy, the Hessian energy can be used for scattered data
interpolation.
One solves the following minimization problem,
for some given interpolation data \(u(x_i) = f_i, \; i=1,...,n\)
\begin{equation}
    \argmin_u E(u) \quad\quad u(x_i) = f_i, \; i=1,...,n
    \;\textrm{.}
\end{equation}
As long as at least three interpolation points are provided, this problem has
a solution.
This is because the null space of the Hessian energy can have at most
all linear functions in it, which is a three-dimensional space, and
the null space of the Laplacian energy with zero Neumann boundary conditions
contains only constant functions, which is a one-dimensional space
\cite{Stein2018}.

The choice of smoothness energy will greatly influence the quality of the
result.
The Laplacian energy with zero Neumann boundary conditions, \(E_{\Delta^2}\),
is a popular method, since it produces smooth, evenly spaced isolines, which
results in natural-looking interpolation and extrapolation.
This is because the gradient of the solution is relatively uniform across the
surface.
As can be seen in \figref{scatteredcomparisonclosedsurface}, our curved Hessian
energy \(E\) reproduces the desirable behavior of the Laplacian energy for
surfaces without boundary.
The implementation of the planar Hessian energy \(E_\Hesst\) for
curved surfaces by \citet{Stein2018} fails to do so:
the distance between the isolines varies greatly, for example on the legs.
The isolines also experience significant bunching at the rump and back of the
horse.

On the other hand, the Laplacian energy is known to produce bias near domain
boundaries due to its low-order boundary conditions:
isolines of solutions bend so they can be perpendicular to the boundary.
This was one of the motivations of \citet{Stein2018}, and thus their planar
Hessian energy minimizes the influence of the boundary by employing natural
boundary conditions that make the function as-linear-as-possible.
\figref{scatteredcomparisonwithholed} shows that our Hessian energy \(E\)
does not show the bias at the boundary that the Laplacian energy does:
this is because it also has as-linear-as-possible natural boundary conditions.

For this application, our Hessian energy \(E\) combines the two worlds
of Laplacian energy and planar Hessian energy to produce a smoothness energy
that is suited for scattered data interpolation on curved surfaces while
unbiased by the presence of boundaries (\figref{teaser},
\figref{vikinghelmet}).
This is helpful if the boundaries of the surface don't have any physical
meaning: perhaps they are the result of a faulty laser scan, or perhaps surface
information is simply not available there.
The Hessian energy's natural boundary conditions make a best guess everywhere
the data is missing by extrapolating the function linearly across the boundary.

\subsection{Data smoothing}

Another popular application for smoothness energies is the eponymous data
smoothing.
This can be used to simply smooth arbitrary data, to denoise noisy data, or
to smooth the surface itself via surface fairing.
One solves the following Helmholtz-like optimization problem:
given an input function \(f\) to be smoothed,
\begin{equation}\label{eq:smoothingproblem}
     u = \argmin_u E(u) + \alpha \int_\Omega (f - u)^2 \;\dx
     \;\textrm{,}
\end{equation}
where the parameter \(\alpha>0\) is a trade-off between the input data and the
smoothness of the output data.

\figref{smoothingresult} shows our Hessian energy \(E\) applied to such a
smoothing problem.
Correctly accounting for curvature by modeling a curved biharmonic equation
using the Laplace-Beltrami operator is important here:
the figure shows that the approach of \citet{Stein2018} causes distortion in
high-curvature regions when smoothing a step function.
In this figure the smoothing parameters are chosen to give visually similar
amounts of smoothing, which means a slightly larger parameter \(\alpha\) for
the method of \citet{Stein2018}.

\begin{figure}
	\includegraphics[width=\linewidth]{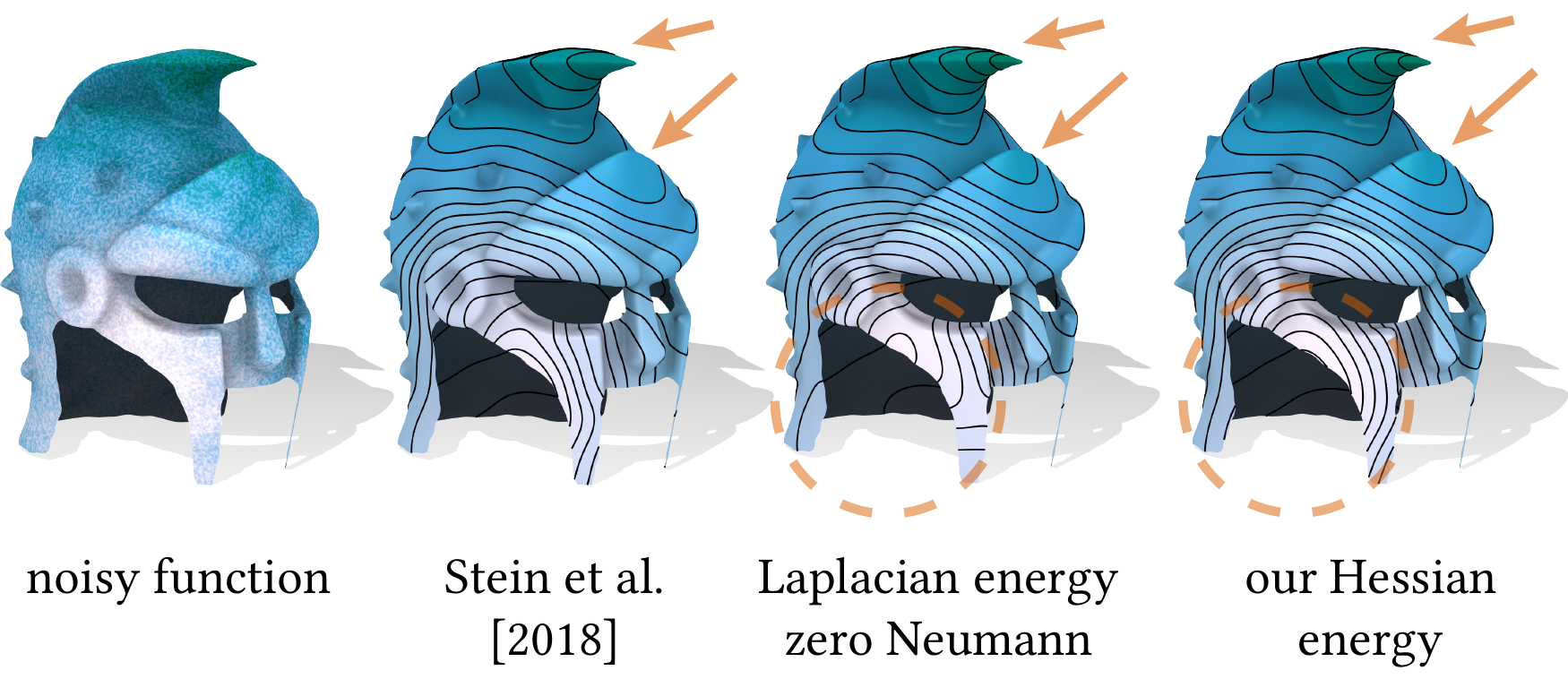}
	\caption{Denoising a function \figloc{far left} via smoothing.
		The Hessian energy \(E\) \figloc{far right} does not show the bias at the
		boundary that the Laplacian energy with zero Neumann boundary conditions
		\(E_{\Delta^2}\) \figloc{center right} does, indicated by the orange circle.
		Away from the boundary, the results for \(E\) and \(E_{\Delta^2}\) agree,
		while the method of \citet{Stein2018} \figloc{center left} differs,
		indicated by the orange arrows.}
	\label{fig:densesmoothingresult}
\end{figure}

It is natural to ask why the fact that minimizers of \(E_\Hesst\)
do not solve the biharmonic equation leads to worse results when smoothing 
the step function of \figref{smoothingresult}, but not for the
smoothing problems solved by \citet[Fig.\ 1, Fig.\ 11, Fig. 13]{Stein2018}.
These examples all smooth very noisy functions with a lot of variation
everywhere on the surface.
The step function is the opposite of that:
the variation is much more sparse.
This allows the error of not accounting for curvature correctly to manifest.
In \figref{densesmoothingresult} such a denoising problem is solved using
the energies \(E_{\Delta^2}\) (with zero Neumann boundary conditions),
\(E_\Hesst\) (with the implementation of \citet{Stein2018}), and \(E\).
It can be clearly seen that \(E_{\Delta^2}\), the Laplacian energy with zero
Neumann boundary conditions, is biased by the boundary, and the isolines
near the boundary are distorted so they can be normal to it.
The denoised solution using the Hessian energy \(E\) does not suffer from this,
and the isolines ignore the boundary.
In regions far away from the boundary it can be observed that the result of
denoising with the Hessian energy \(E\) matches the Laplacian energy with zero
Neumann boundary conditions \(E_{\Delta^2}\), while the planar Hessian energy
\(E_\Hesst\) differs.

\begin{figure}[t]
	\includegraphics[width=\linewidth]{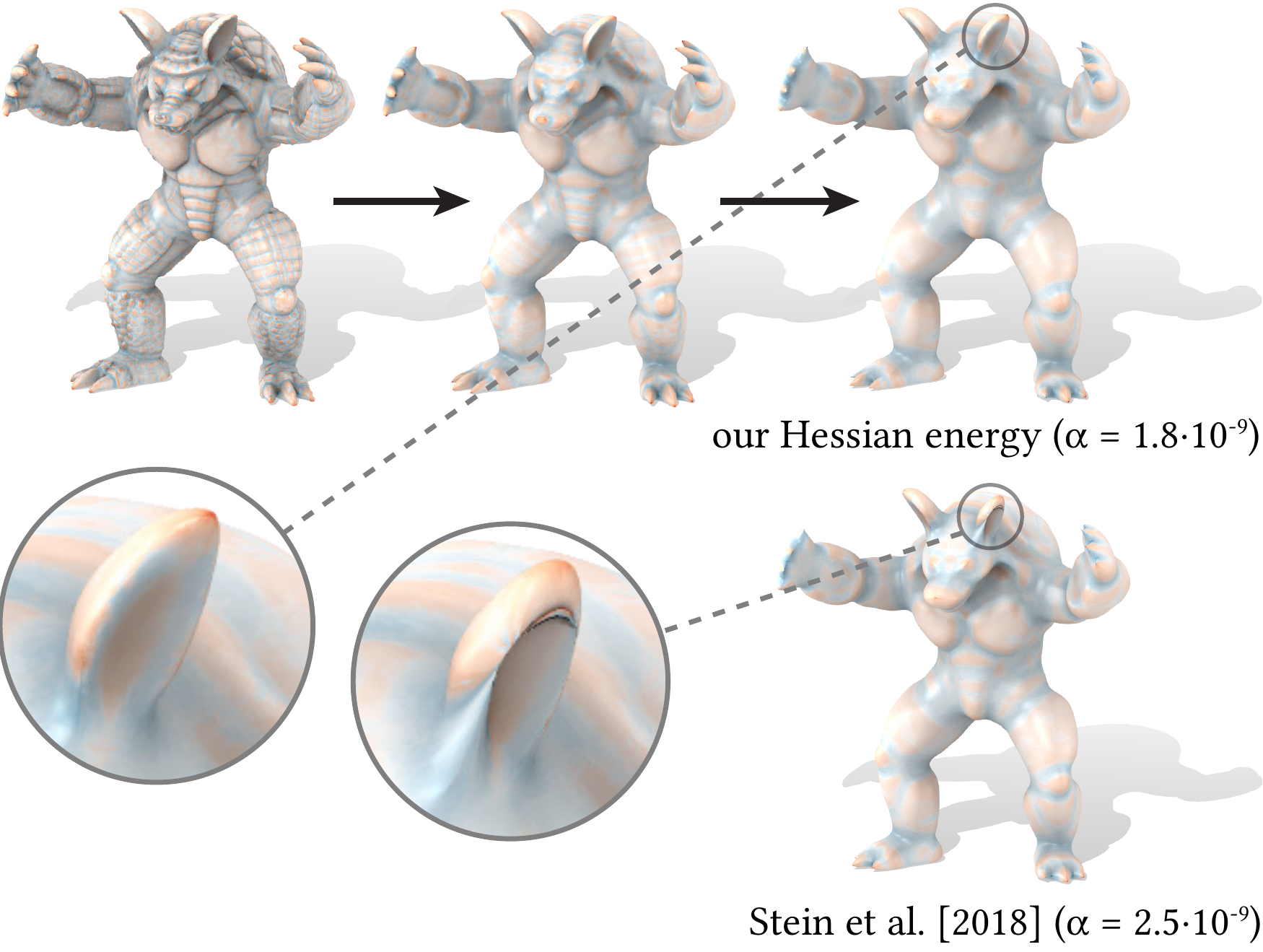}
	\caption{Smoothing flow for an armadillo.
		The surfaces are colored by angle defect.
		Each step of our Hessian energy \(E\) \figloc{top} leads to a smoother result.
		Smoothing with \citet{Stein2018} \figloc{bottom} can lead to artifacts
		in regions with curvature, such as the highlighted ears.
		\\
		The smoothing parameter \(\alpha\) was chosen to produce a similar amount of
		smoothing in both methods.
		Three smoothing steps were computed.}
	\label{fig:surfacefairingresult}
\end{figure}

The smoothing problem can also be used to smooth the geometry of
the surface itself if the input data \(f\) from \equref{smoothingproblem} is
the vertex positions in each coordinate, and the output data \(u\) is the new
vertex positions.
If such a smoothing operation is applied repeatedly, one has a smoothing flow.
\figref{surfacefairingresult} shows our Hessian energy \(E\) applied to such a
problem.
While the method of \citet{Stein2018} can lead to some artifacts due to
not accounting for curvature, this does not happen with our curved Hessian
energy \(E\).

\section{Conclusion}
\label{sec:conclusion}

In this work we have introduced a smoothing energy for curved surfaces, the
Hessian energy.
Its minimizers solve the biharmonic equation, and it exhibits the 
as-linear-as-possible natural boundary conditions in the curved setting
that the planar Hessian energy of \citet{Stein2018} exhibits in the flat
setting.
This Hessian energy can be used in many applications where smoothness energies
are required, these smoothness energies should be unbiased by the
boundary, and it is crucial that the minimizers of the energy solve the
biharmonic equation.

\subsection{Limitations}

We have no numerical analysis proof for the convergence of our 
discretization method.
We also do not provide any theoretical analysis of the spectrum of our 
discrete operator.
Both are needed to make this discretization reliable, and to improve
understanding of the method, where it works, and where it does not.

\subsection{Future work}

One interesting avenue for future work is to explore alternate discretizations.
Higher-order versions of Crouzeix-Raviart basis functions, such as cubic or
quintic basis functions, would be an interesting potential improvement.
Alternatively, instead of choosing the intermediate variable \(w = \dd v\) for
the mixed formulation as in \equref{aconstrainedproblem}, a discretization where
\(w = \nabla\dd v\) sounds very promising.
This would more closely mirror the mixed formulation of \citet{Stein2018}.
The CROF approach can be used to define a basis for tensors in the same way
as is done for vectors in \secref{crof}, based on the parallel and the
perpendicular vector at each edge.
Using other finite elements to discretize
the space of one-forms could also produce new methods.
Moreover, future work could explore discretizations of the smooth energy
on other surface representations beyond triangle meshes.

A rich source of future work is the numerical analysis of our method.
We do not have any proof of convergence, or a solid mathematical analysis
of the spectrum of our operator, and while the experiments in
\secref{performance} provide some evidence for problems that can be solved with
our discretization of \(E\), a thorough numerical analysis treatment of our
discretization would be valuable to exactly identify the strengths and
weaknesses of our method.
Our Crouzeix-Raviart discretization is a potential candidate for spurious
modes, since the finite element is non-conforming, even though we have not
observed them in practice.
The method of \citet{English2008} is an example of a Crouzeix-Raviart
discretization that works for many cases, but where specific triangle
configurations exist that lead to spurious modes
\cite[Section 4.4.2]{Quaglino2012}.
The properties of minimizers of the discrete energies also warrant further
investigation:
it is unclear which properties of smooth minimizers they actually inherit,
and which properties only hold in the limit.

Another interesting direction for future work is to consider additional
applications.
Smoothness energies have many uses, and if such an application has to be
unbiased by the boundary even on heavily curved surfaces, our Hessian energy
\(E\) is a powerful tool.
Applications could include animation \cite{Jacobson2011}, distance computation
\cite{Crane2013Geodesics}, and more.

Moreover, our simple Crouzeix-Raviart discretization of the one-form
Dirichlet energy containing covariant derivatives from \secref{crof}
offers an interesting approach to discretize the vector Dirichlet energy
in a wide variety of applications.
Potential applications include vector field design \cite{Knoppel2013},
parallel transport of vectors \cite{Sharp2018}, and many more
\cite{Azencot2015,Liu2016,Corman2019}.

\section*{Acknowledgements}

We thank the libigl data repository \cite{libigldata} (horse, camel, cross
lilium, hand, puppet head by Cosmic blobs), the Stanford 3D Scanning Repository
\cite{stanfordrepository} (armadillo), \citet{keenansmodels} (rubber duck,
man-bridge, spot the cow, Nefertiti by Nora Al-Badri and Jan Nikolai Nelles),
\citet{alecthesis} (various cartoons in \figref{forwardproblems}),
\citet{planeobj} (plane), \citet{towerobj} (tower), and
\citet{helmetobj} (helmet) for some of the
meshes used in this work.

This work is funded in part by the National Science Foundation Awards
CCF-17-17268 and  IIS-17-17178.
This research is funded in part by NSERC Discovery (RGPIN2017-05235,
RGPAS-2017-507938), the Canada Research Chairs Program, the Fields Centre for
Quantitative Analysis and Modelling and gifts by Adobe Systems, Autodesk and
MESH Inc.
This work is partially supported by the DFG project 282535003:
Geometric curvature functionals: energy landscape and discrete methods.

We thank Anne Fleming, Henrique Maia and Peter Chen for proofreading.
\\\quad\\

\afterpage{\clearpage}
\begin{figure*}
    \includegraphics[width=\linewidth]{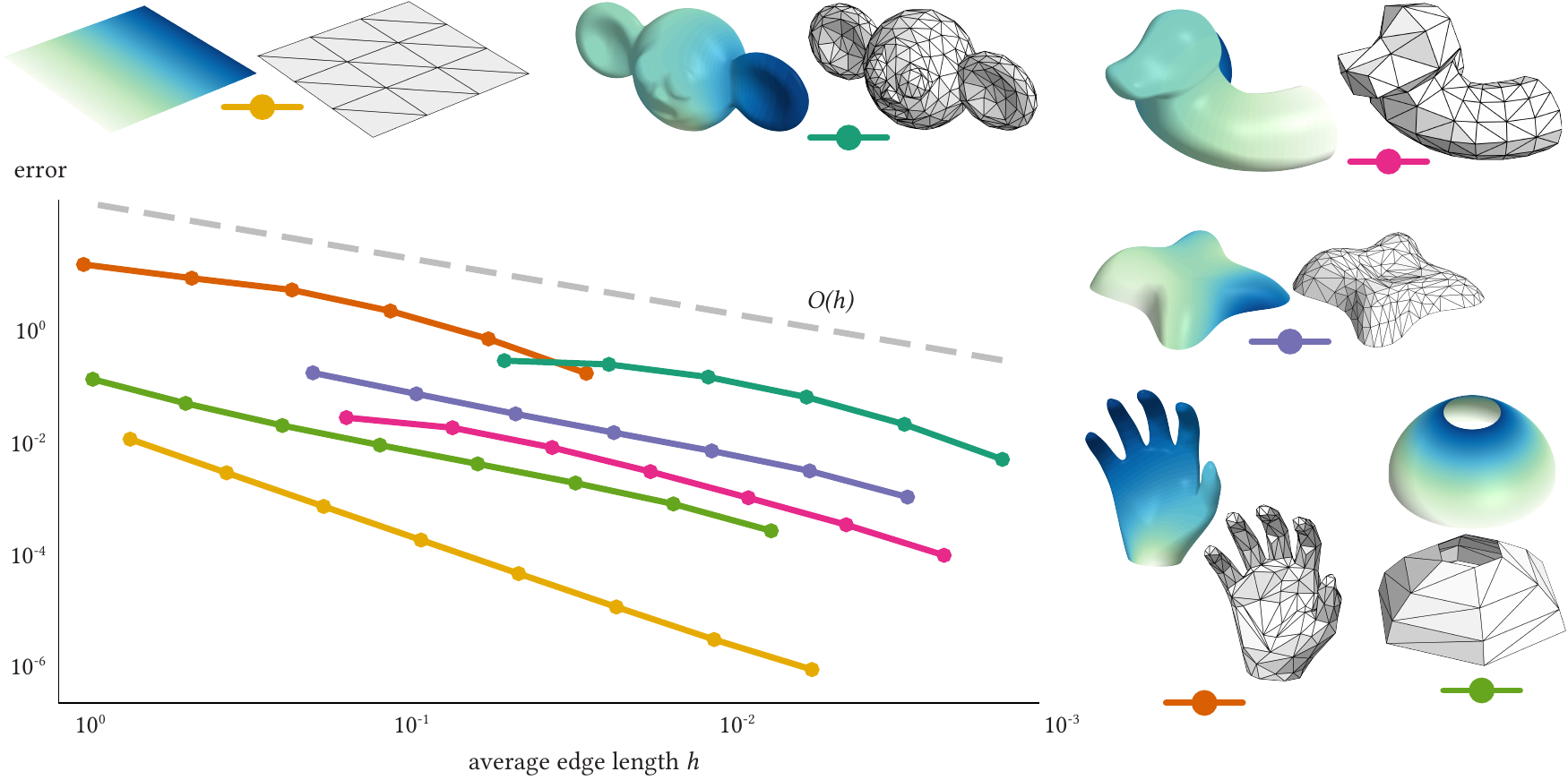}
    \caption{Error plot for six different boundary value problems.
        The minimizer of the Hessian energy \(E\) discretized with our discretization
        is compared to a high-resolution solution with the same discretization.
        Refinement happens via loop subdivision with various types of fixed boundary.
        The high-resolution solution as well as the wireframe of the
        lowest-resolution mesh are displayed for each problem.}
    \label{fig:manyconvergenceplots}
\end{figure*}

\begin{figure*}[b]
    \includegraphics[width=\linewidth]{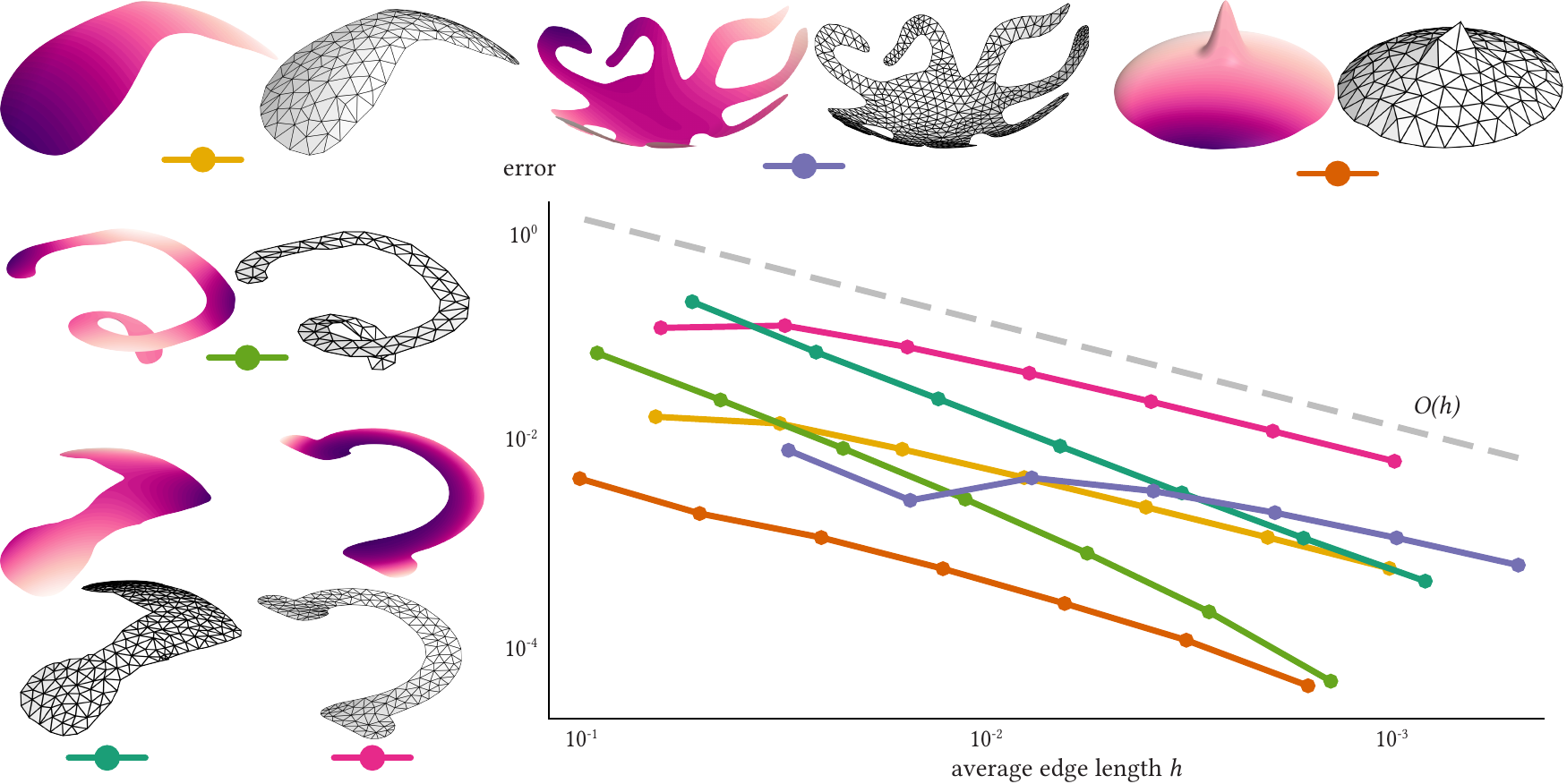}
    \caption{Error plot for six different forward problems.
The domains are curved surfaces of the form \(\left(x, y, z(x,y) \right)
        \in \Rthree\), so the integrand of the Hessian energy
        can be exactly computed pointwise using the properties of Monge patches
        \cite{mongepatches}.
        Quadrature is then used to compute the exact value of \(E(f)\).
        The high-resolution function \(f\) as well as the wireframe of the
        lowest-resolution mesh are displayed for each problem.
        Refinement happens via loop subdivision, and then projection to the
        given smooth surface.}
    \label{fig:forwardproblems}
\end{figure*}
\afterpage{\clearpage}
 
\bibliographystyle{ACM-Reference-Format}
\bibliography{curvature-hessian}

\appendix

\section*{Appendix}

\section{Implementation}
\label{app:implementation}

\begin{figure}[t]
 \includegraphics[width=\linewidth]{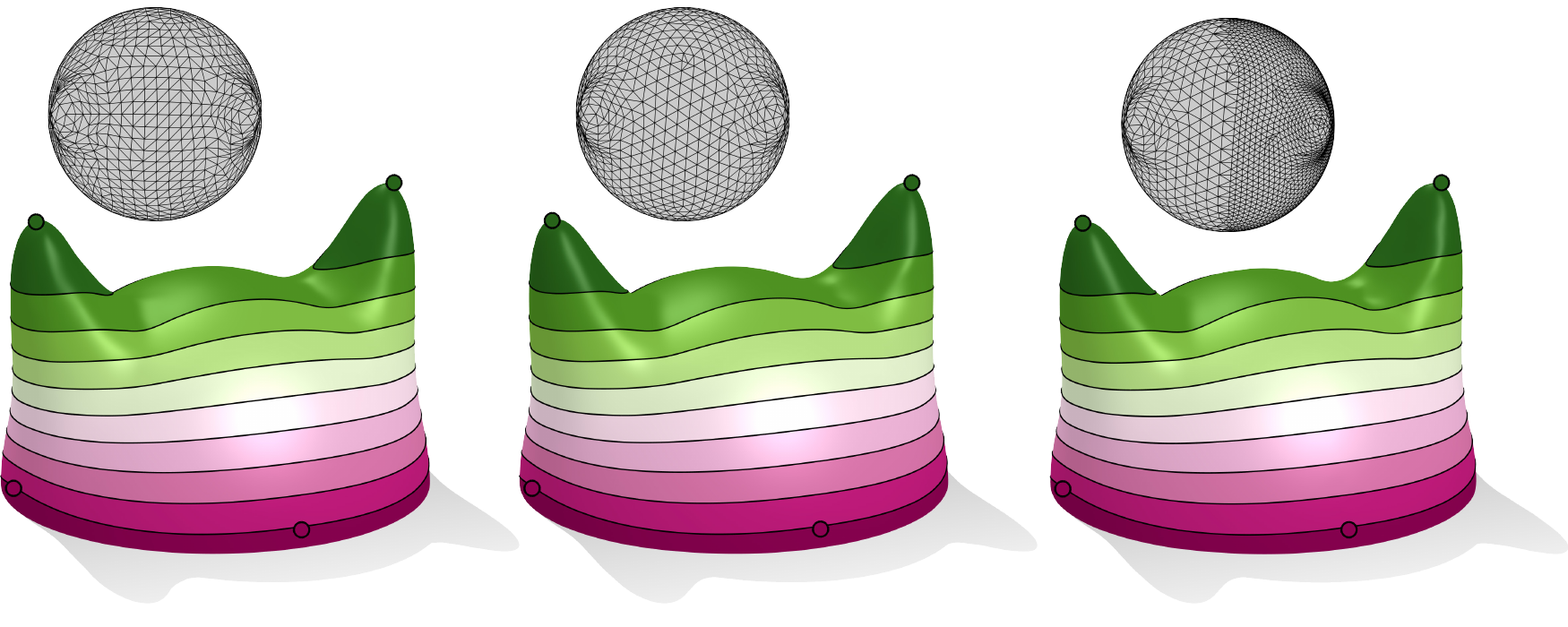}
 \caption{The same scattered data interpolation problem solved on different
 meshes for surfaces similar to the one from \figref{vikinghelmet} using the
 Hessian energy \(E\).
 The results are very similar.
 The wireframe shows each of the meshes before further refinement through
 loop subdivision with fixed boundary.}
 \label{fig:meshingdependence}
\end{figure}

The entries for each of the matrices defined in \secref{computingthehessian}
needed to construct the system matrices used in
\equref{simplifieddiscreteenergy} are as follows.
Let \(e\) be an oriented edge from the vertex \(i\) to \(j\).
The two triangles adjacent to \(e\) are \(T_{ijk}\) and \(T_{jil}\), and \(f\)
is an oriented edge from the vertex \(k\) to \(i\).
The entries of the symmetric CROF vector Dirichlet matrix \(L\) on the
triangle \(T_{ijk}\) are
\begin{equation}\begin{split}
    L_{e^\parallel, e^\parallel}^{ijk} = L_{e^\perp, e^\perp}^{ijk}
    &= \frac{2}{A_{ijk}} \\
    L_{e^\parallel, e^\perp}^{ijk} &= 0 \\
    L_{e^\parallel, f^\parallel}^{ijk} = L_{e^\perp, f^\perp}^{ijk}
    &= \frac{2}{A_{ijk}} \cos^2 \theta_i^{ijk} \\
    L_{e^\perp, f^\parallel}^{ijk} = -L_{e^\parallel, f^\perp}^{ijk}
    &=  \frac{2}{l_{ij}l_{ki}} \cos \theta_i^{ijk}
    \;\textrm{,}
\end{split}\end{equation}
where \(A_{ijk}\) is the double area of the triangle \(T_{ijk}\),
\(\theta_i^{ijk}\) is the angle in the triangle \(T_{ijk}\) at the vertex
\(i\), and \(l_{ij}\) is the length of the edge from vertex \(i\) to \(j\).
If one of the edges has reversed orientation in the triangle \(T_{ijk}\) with
respect to its global orientation, its off-diagonal entries get multiplied by
\(-1\).
These are only the terms for the triangle \(T_{ijk}\).
One must add the terms for all triangles and all pairs of edges in that
triangle to compute the full matrix \(L\).
We suggest looping through all triangles, and adding the terms for each
triangle to the respective entries of the matrix corresponding to the edges.
This can easily be parallelized with a \texttt{parallel\_for} loop.

The entries of the diagonal CROF mass matrix \(M\) on the triangle \(T_{ijk}\)
are
\begin{equation}\begin{split}
    M_{e^\parallel, e^\parallel}^{ijk} = M_{e^\perp, e^\perp}
    &= \frac{A_{ijk}}{6 l_{ij}^2}
    \;\textrm{.}
\end{split}\end{equation}

The entries of the differential matrix \(D\) on the triangle \(T_{ijk}\) for each
edge \(e\) are
\begin{equation}\begin{split}
    -D_{i, e^\parallel}^{ijk} = D_{j, e^\parallel}^{ijk}
    &= \frac{A_{ijk}}{6 l_{ij}^2} \\
    D_{k, e^\parallel}^{ijk} &= 0 \\
    D_{i, e^\perp}^{ijk} &= - \frac{l_{jk}}{6 l_{ij}}\cos\theta_j^{ijk} \\
    D_{j, e^\perp}^{ijk} &= - \frac{l_{ki}}{6 l_{ij}}\cos\theta_i^{ijk} \\
    D_{k, e^\perp}^{ijk} &= \frac{1}{6}
    \;\textrm{,}
\end{split}\end{equation}
where \(i\) is the vertex at the tail of the edge \(e\), and \(j\) is at
its tip.
If one of the edges has reversed orientation in the triangle \(T_{ijk}\) with
respect to its global orientation, its entries get multiplied by
\(-1\).

\begin{figure}[t]
 \includegraphics[width=\linewidth]{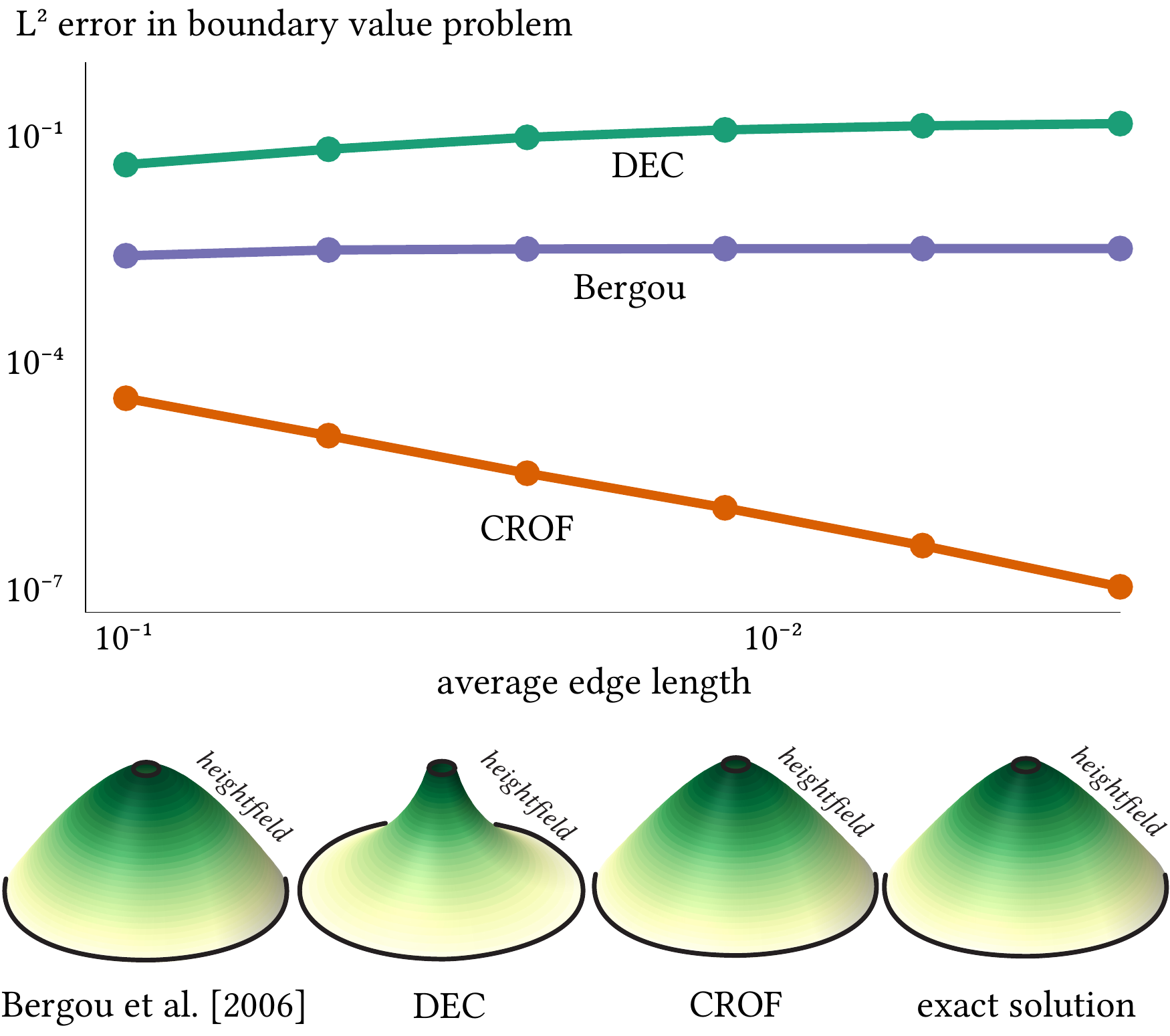}
 \caption{A comparison of the CROF Hessian, the DEC Hessian
 (as of \citet[(20)]{Stein2018}, described by \citet{Fisher2007} and
 implemented by \citet{Wang2015}), and the Bergou Hessian
 (as of \citet[(21)]{Stein2018}, described by \citet{Bergou2006} and
 implemented by \citet{Wang2017}) in green.
 The two non-CROF Hessians fail to match the exact solution on the
 annulus, even though the method of \citet{Bergou2006} looks visually 
 similar.}
 \label{fig:deccomparisons}
\end{figure}

The entries of the curvature correction matrix \(K\) on the triangle
\(T_{ijk}\) are
\begin{equation}\begin{split}
    K_{e^\parallel, e^\parallel}^{ijk} = K_{e^\perp, e^\perp}^{ijk}
    &= \frac{1}{l_{ij}^2} \left( \frac{\theta_i^{ijk}}{s_i} \kappa_i
    + \frac{\theta_j^{ijk}}{s_j}\kappa_j
    + \frac{\theta_k^{ijk}}{s_k} \kappa_k \right) \\
    K_{e^\parallel, e^\perp}^{ijk} &= 0 \\
    K_{e^\parallel, f^\parallel}^{ijk} = K_{e^\perp, f^\perp}^{ijk}
    &= \frac{\cos\theta_i^{ijk}}{l_{ij} l_{ki}}
    \left( \frac{\theta_j^{ijk}}{s_j}\kappa_j
    + \frac{\theta_k^{ijk}}{s_k} \kappa_k
    - \frac{\theta_i^{ijk}}{s_i} \kappa_i \right) \\
    - K_{e^\parallel, f^\perp} = K_{e^\perp, f^\parallel}
    &= \frac{\sin\theta_i^{ijk}}{l_{ij} l_{ki}}
    \left( \frac{\theta_j^{ijk}}{s_j}\kappa_j
    + \frac{\theta_k^{ijk}}{s_k} \kappa_k
    - \frac{\theta_i^{ijk}}{s_i} \kappa_i \right)
    \;\textrm{,}
\end{split}\end{equation}
where \(\kappa_v\) is the angle defect at the vertex \(v\) and \(s_v\) is the
angle sum at the vertex \(v\).
If one of the edges has reversed orientation in the triangle \(T_{ijk}\) with
respect to its global orientation, its off-diagonal entries get multiplied by
\(-1\).

\section{Additional experiments}
\label{app:additionalexperiments}

\figref{manyconvergenceplots} features a series of convergence experiments
that shows the convergence of a boundary value problem on a variety of meshes
to the highest-resolution solutions.
In \figref{forwardproblems}, a series of forward problems is solved,
where the Hessian energy of a function is measured on a curved
surface, and because both the function and the surface embedding are known,
the exact solution is also known.
This is used to measure the error.
In both these examples, convergence of the order of the average edge length
is observed.

\figref{meshingdependence} shows that for different meshings of the same
surface, very similar results are achieved, and the method is thus robust
to remeshing.
In \figref{deccomparisons} our CROF implementation of the Hessian energy is
compared with various Hessian energies discussed by \citet{Stein2018} in
the flat annulus setting, where the exact solution is known.

\end{document}